%% file: paper.tex
\documentclass[iop]{emulateapj}

\shortauthors{Reiners et al.}

\begin{document}


\title{A catalogue of rotation and activity in early-M stars}


\author{Ansgar Reiners and Nandan Joshi}
\affil{Institut f\"ur Astrophysik G\"ottingen, Physik Fakult\"at,
  Friedrich-Hund-Platz 1, D-37077 G\"ottingen}
\email{Ansgar.Reiners@phys.uni-goettingen.de}

\and

\author{Bertrand Goldman}
\affil{Max-Planck-Institut f\"ur Astronomy, Heidelberg}



\begin{abstract}
  We present a catalogue of rotation and chromospheric activity in a sample of
  334 M dwarfs of spectral types M0--M4.5 populating the parameter space
  around the boundary to full convection. We obtained high-resolution optical
  spectra for 206 targets and determined projected rotational velocity,
  $v\,\sin{i}$, and H$\alpha$ emission.  The data are combined with
  measurements of $v\,\sin{i}$ in field stars of the same spectral type from
  the literature.  Our sample adds 157 new rotation measurements to the
  existing literature and almost doubles the sample of available
  $v\,\sin{i}$. The final sample provides a statistically meaningful picture
  of rotation and activity at the transition to full convection in the solar
  neighborhood. We confirm the steep rise in the fraction of active stars at
  the transition to full convection known from earlier work. In addition, we
  see a clear rise in rotational velocity in the same stars. In very few
  stars, no chromospheric activity but a detection of rotational broadening
  was reported. We argue that all of them are probably spurious detections; we
  conclude that in our sample all significantly rotating stars are active, and
  all active stars are significantly rotating. The rotation-activity relation
  is valid in partially and in fully convective stars. Thus, we do not observe
  any evidence for a transition from a rotationally dominated dynamo in
  partially convective stars to a rotation-independent turbulent dynamo in
  fully convective stars; turbulent dynamos in fully convective stars of
  spectral types around M4 are still driven by rotation. Finally, we compare
  projected rotational velocities of 33 stars to rotational periods derived
  from photometry in the literature and determine inclinations for a few of
  them.
\end{abstract}



\section{Introduction}

M-dwarfs constitute the majority of stars in the solar
neighborhood. They are intrinsically faint because they are cooler and
smaller than all other stars, and their physical properties span more
than a factor of five in mass and radius from the coolest late-M-type
stars and young brown dwarfs with less than a tenth of solar mass to
the most massive M dwarfs with more than half a solar mass. Because of
their faintness, detailed spectroscopic investigation is more
observationally demanding than the analysis of brighter objects, but
the sheer number of M stars renders them an excellent statistical
sample for understanding properties of stellar physics and
evolution. Many M dwarfs show substantial magnetic activity resulting
in chromospheric and coronal heating, observed in various indicators
across the whole stellar spectrum. It is often argued that M dwarfs
are places of violent energy outbursts unsuitable for the existence of
life. However, a large fraction of M dwarfs show little signs of
magnetic activity, and M dwarfs have become a prime target for the
search for Earth-like extrasolar planets.

The spectral type regime of early- to mid-type main sequence M stars
coincides with the mass regime where the transition from partial
convection to full convection occurs
\citep[]{Chabrier:1997p2277}. Partially convective stars, or
solar-type stars, are believed to generate at least parts of their
magnetic fields through a global dynamo residing at the interface
between the radiative core and the convective envelope. Rotational
shear at this interface can amplify magnetic fields and sustain a
cyclic magnetic dynamo \citep[]{Parker:1993p2881,
  Ossendrijver:2003p2903}. Field lines end up rising to the surface of
the star and become visible in the form of starspots. Rotation,
therefore, is the main driving force behind chromospheric and coronal
activity.  However, at $M \sim 0.35\,$M$_{\sun}$ stars are believed to
become fully convective and no interface layer exists anymore in their
interior. The stars also suffer significant structural changes leading
to dramatic differences in mass and radius while effective temperature
(spectral type) only changes little. Therefore, radius and mass are
strongly related to spectral subtype around the boundary to complete
convection, which can explain the observed change in braking
efficiency and activity lifetimes in this mass regime \citep{ReiMo11}.

Despite the changes in stellar structure, strong magnetic activity also
appears in very-low mass stars that are fully convective
\citep[]{Hawley:1996p1924, West:2004p78}. Observations of magnetically
sensitive molecular lines \citep[]{Reiners:2007p19, 2011MNRAS.tmp.1579S},
Zeeman Doppler imaging of M-stars \citep[]{Donati:2008p2516, Morin:2008p2170}
as well as numerical simulations of magnetic field generation in fully
convective stars \citep[]{Browning:2008p2191} agree that strong magnetic
fields exist across the full range of M-type dwarfs.

Rotation plays a crucial role in all scenarios of magnetic field
generation. Observations of activity in solar-type stars show a direct
connection between rotation and activity, the so-called
rotation-activity relation \citep[]{Noyes:1984p48, Delfosse:1998p79,
  Pizzolato:2003p74}. Activity grows stronger with increasing
rotational velocity and saturates at a threshold velocity that depends
on the mass of the star \citep[]{Pizzolato:2003p74}. The Rossby number
$Ro = P/\tau_{\rm conv}$, with $P$ the rotation period and $\tau_{\rm
  conv}$ the convective overturn time, is often used as a unifying
scale of actvity; activity saturates around $Ro = 0.1$. A
saturation-type rotation-activity relation in M dwarfs is observed for
late-M spectral types \citep[later than M5;][]{Mohanty:2003p83,
  2010ApJ...710..924R}, and less well studied also in early-M type
stars \citep[]{Delfosse:1998p79, Reiners:2007p19}.

In this paper, we concentrate on early- to mid-M dwarfs including the
transition from partial to complete convection. We include in our catalogue
only stars of spectral type M0--M4.5. We took new observations that we combine
with data from the literature. Our sample selection is explained in
Section\,\ref{sect:Sample}, analysis methods are described in
Section\,\ref{sec:analysis} and discussed in Section\,\ref{sec:results}.


\section{Observations and Data reduction}
\label{sect:Sample}

\subsection{New observations}

The goal of our project is to construct a statistically meaningful,
yet neither complete nor unbiased, sample of early-M field dwarf
spectra. Literature available at the time of observations containing
considerable samples of high spectral-resolution analysis of field-M
dwarf rotation velocities were \citet{Marcy:1992p2101} and
\citet{Delfosse:1998p79}. During the course of our project,
\citet{Jenkins:2009p2650} and \citet{Browning:2010p2615} added more
stars to this list, and a few other rotational velocities were
presented in \citet{Reiners:2007p18} and \citet{Reiners:2007p19}. We
did not exclude young or halo stars from the sample in order to
achieve a representative picture of the stars in the solar
neighborhood.

Observations for this project were carried out at the spectrographs
FOCES (CAHA, Calar Alto) in 2005, and FEROS (ESO, La Silla) in 2006
using time allocated through the the Max-Planck Institute for
Astronomy (MPIA) at both observatories. In total, we have obtained
spectra of 239 M0--M4.5 stars.

\subsection{Data reduction}

The FOCES echelle spectrograph at the 2.2\,m Telescope at CAHA, Calar
Alto, was operated at spectral resolving power of approximately
40,000. Observations carried out with FEROS at the ESO/MPG 2.2\,m
telescope at La Silla have a spectral resolution of approximately
48,000. Typical exposure times are between a few minutes and one hour
per star for FEROS and up to two hours with FOCES. Individual
exposures are always shorter than 30\,min to avoid crowding with
cosmic rays. The resulting signal-to-noise ratios are typically around
50, further analysis uncertainties are discussed in
Section\,\ref{sec:analysis}.

Data reduction followed standard procedures including bias subtraction,
flat fielding, and wavelength calibration provided by ThAr or ThArNe lamps. For
the FEROS spectra, reduction was done with the dedicated pipeline based on the
MIDAS context, the FEROS Data Reduction System (DRS).

FOCES data was reduced using standard reduction procedures implemented
in ESO-MIDAS. To avoid over- or underexposure for different
\'{e}chelle orders, three flat fields were taken with different
exposure times, which is the usual procedure for FOCES data
reduction. Orders were grouped in three sets each corresponding to
different flat field exposure times, longest time corresponding to red
and shortest time to blue regions in the spectrum. All three flat
fields were merged to create a master flat field. The raw object
spectra were freed from cosmic ray contamination and, after bias
subtraction, flat-fielded with the master flat field. The wavelength
scale was calibrated for each night using ThAr lamp
calibrations. Scattered light was removed using a standard background
subtraction routine in ESO-MIDAS. Since FOCES is a fiber-fed
spectrograph, the object spectra were first extracted and then divided
by the order-extracted master flat field.

\section{The Catalogue}

In order to provide a comprehensive collection of currently available
information on rotation and activity in early-M dwarfs, we created a
catalogue of results from high-resolution spectra merging our results
with catalogues published earlier and available in the literature. We
tried to select work that used data of quality similar to ours and
that were not selected according to physical parameters of the
targets, i.e., we considered only work collecting data from selections
of early-M field dwarfs. We limited our study to literature values of
the objects in the same spectral range, M0.0--M4.5. Catalogues of this
kind are \citet[]{Marcy:1992p2101}, \citet[]{Delfosse:1998p79},
\citet[]{Reiners:2007p18}, \citet[]{Reiners:2007p19},
\citet[]{Jenkins:2009p2650}, and \citet[]{Browning:2010p2615}. From
\citet{Jenkins:2009p2650}, we used only their own observations given
in their Table\,1 since the collection in their Table~3 is rather
inhomogeneous and contains work focusing on young stars. We did not
include results from the SACY sample \citep{2006A&A...460..695T}
either since we focus on nearby field stars while the SACY targets
form a specifically selected sample of young stars.  All work
considered provide information on $v\,\sin{i}$, but unfortunately
quantitative information on H$\alpha$ emission is not available in all
cases.

\begin{figure}
  \plotone{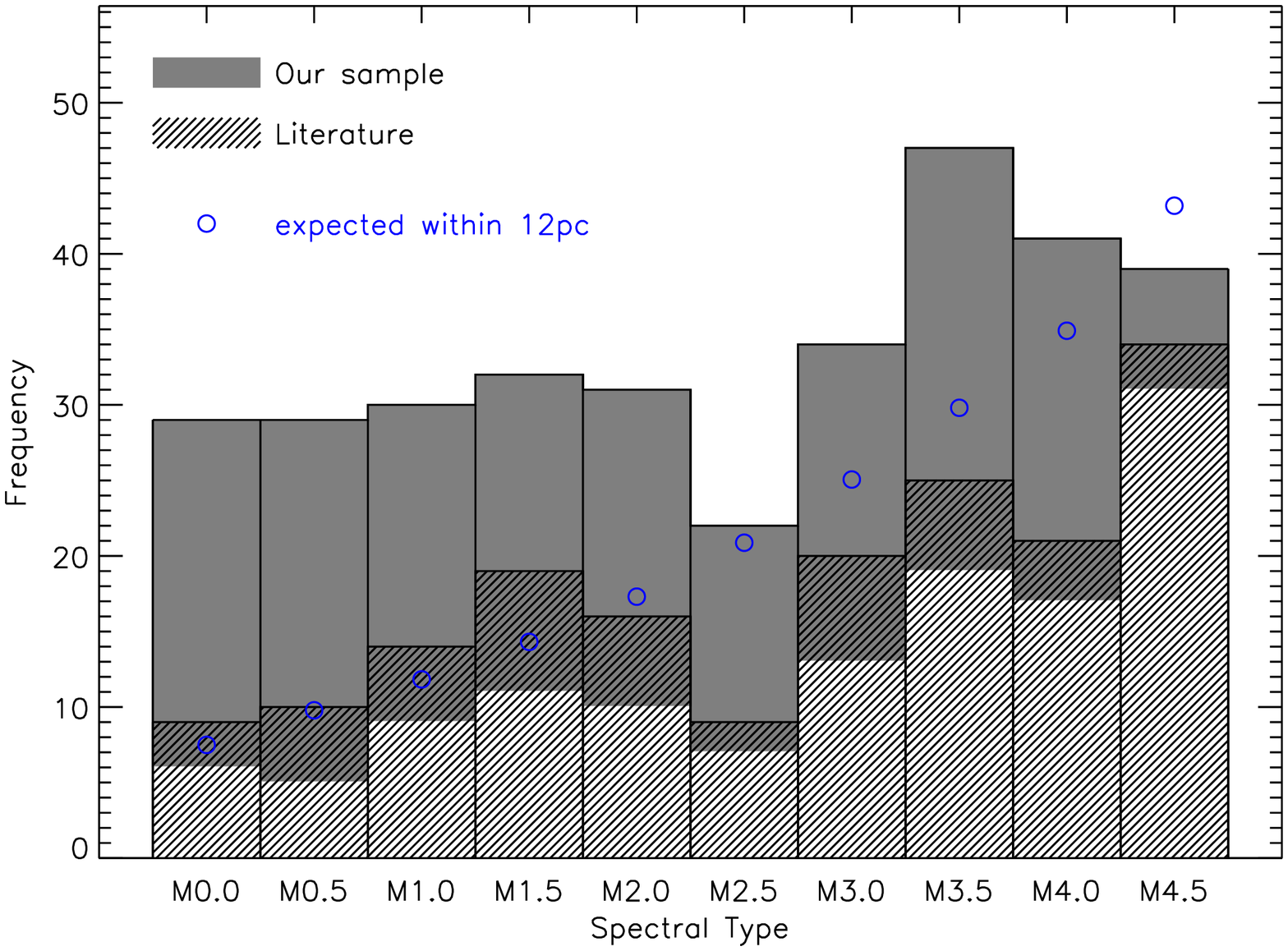}
  \caption{\label{plot:histo_cs}Spectral type distribution of our
    catalogue. The fraction of stars observed within the course of
    this project is shown in grey, stars taken from the literature are
    shown as hatched histogram. Blue circles show expected numbers of
    stars per spectral bin contained within 12\,pc according to the
    mass function from \citet{2010AJ....139.2679B}.}
\end{figure}

The distribution of stars as a function of spectral type is shown in
Fig.\,\ref{plot:histo_cs}. In this sample, known spectroscopic binaries are
already removed (Section\,\ref{ssec:spec_bin}). In total, the sample consists
of 334 M dwarfs of spectral types M0.0--M4.5. Spectra of 206 targets were
taken during the course of this project, several of them were already
available in the literature. Our observations add 157 new measurements of
$v\,\sin{i}$ to the full catalogue. 

The distribution of stars in our sample does not follow the
distribution of stars in the solar neighborhood; in comparison to the
distribution of M dwarfs, early-type stars are overrepresented in our
sample. We can compare the distribution of stars in our sample to the
mass function in the field reported by \citet{2010AJ....139.2679B}. We
use their system mass function parametrization in the lognormal
form. In order to compare our distribution of spectral types to the
mass function, we determine the mass of each spectral type bin and
calculate the expected number of stars per bin. We assume the relation
between spectral type and effective temperature according to the
relation given in \cite{Kenyon:1995p2462} and derive mass from
temperature following the models at 1~Gyr given in
\citet{1998A&A...337..403B}. For simplicity, we assume constant bin
sizes in mass for our spectral type bins, which is a reasonable
approximation in our spectral type regime. For each spectral bin, we
derive its range in $\log{M}$ and determine the number of stars within
a 12\,pc volume as expected from the mass function given in
\citet{2010AJ....139.2679B}. The result is shown as blue circles in
Fig.\,\ref{plot:histo_cs}. A volume-complete sample would include more
mid-M stars per spectral bin than early-M stars. If our sample in each
bin only contained the nearest stars, our sample would cover the
volume out to $\approx 19$\,pc for M0 stars and slightly less than
12\,pc for M4.5 stars. Although our sample cannot be considered
complete out to any given volume, this shows how representative the
sample is for the local Galaxy for the different stars contained; with
respect to the local mass distribution of stars, early-M stars are
overrepresented with respect to mid-M stars.

Our sample probes the local population of M stars. The sample of
early-type stars is drawn from a population that extends up to two
times further than our mid-M type stars. Since we do not expect
significant differences in the properties of rotation and activity as
a function of distance (between 10 and 20\,pc), we do not expect that
this influences the results respective to a volume-complete local
sample. In any case, our results can be interpreted as representative
for typical magnitude-limited surveys (as for example planet hunting
missions will be).

\subsection{$v\,\sin{i}$}

Although the methodologies determining $v\,\sin i$ employed in the different
studies are basically identical, there are some differences in their
implementation. The basic idea is to compare the spectrum of a known, slowly
rotating star (the template spectrum) with a spectrum of the science target in
which the value of $v\sin i$ is to be determined. The template spectrum is
convolved with rotational broadening profiles according to a set of different
velocities using the scheme described in \citet[]{2005oasp.book.....G}.

There are in principle two methods used to determine $v\sin i$. The
first method is to directly compare a chosen set of individual
spectral lines to artificially broadened template spectra. The value
of $v\sin i$ is the one providing the best fit. This method is prone
to systematic uncertainties induced by a mismatch between the template
and science objects' spectra.  The line profiles for comparison must
be selected very carefully and systematic differences may occur if
different sets of lines are used. This method determining $v\sin i$
was employed by \citet[]{Marcy:1992p2101}, \citet[]{Reiners:2007p18},
\citet[]{Reiners:2007p19}, and \citet[]{Jenkins:2009p2650}. While
\citet[]{Marcy:1992p2101} and \citet[]{Jenkins:2009p2650} used atomic
lines at optical wavelength ranges where blending is a serious issue,
\citet[]{Reiners:2007p18} and \citet[]{Reiners:2007p19} employed lines
of molecular FeH that are relatively free of blends and embedded in a
rather well defined continuum.

The second method to derive values of $v\,\sin{i}$ is the so-called
cross-correlation technique \citep[]{Tonry:1979p1685, 2000ApJ...538..363B}.
Here, a slowly rotating template star is identified and its spectrum is
cross-correlated with a series of the same template spectrum that is
artificially broadened according to different rotational velocities. The
widths of the correlation functions provide a calibrated measure of the value
of $v\,\sin{i}$. Then, the target spectrum is cross-correlated with the
non-broadened template spectrum and the width of the resulting correlation
peak is converted into $v\,\sin{i}$ according to the calibration.  Spectra are
usually divided into several sections (e.g., spectral orders), and
cross-correlation functions of individual orders may be averaged, or the
median of individually derived $v\,\sin{i}$ values from different orders can
be used. The latter procedure may also provide an estimate of the
uncertainty. This cross-correlation method was employed in the analysis (see
Section\,\ref{ssec:vsini_det}) of our new observations and in the work of
\citet[]{Delfosse:1998p79} and \citet[]{Browning:2010p2615}.

\subsection{Chromospheric activity}

We collected measurements of projected rotational velocities from the
literature. Unfortunately, not all literature also provide activity
measurements together with rotation, or they provide only information
on whether H$\alpha$ emission is detected or not, but not the value of
equivalent width.

\citet{Delfosse:1998p79} and \citet{Browning:2010p2615} include values
of $\log{\rm L_{\rm H_{\alpha}}/L_{\rm bol}}$ in their tables, and we
included their results in our catalogue. \citet{Marcy:1992p2101} do
not provide information on H$\alpha$
emission. \citet{Jenkins:2009p2650} provide only information on
whether H$\alpha$ is detected, and whether they find H$\alpha$ in
absorption in their spectra. If no H$\alpha$ was detected at all, we
calculate upper limits for the stars assuming similar detection
thresholds as for our data (see Section\,\ref{ssec:halpha_det})
because the data quality is similar. In cases where
\citet{Jenkins:2009p2650} found H$\alpha$ in absorption, we treated
these stars as inactive but mark them in our table. Although the
existence of H$\alpha$ in absorption may be an indicator of weak
activity \citep[]{Cram:1979p2487}, we classified these stars as
inactive because all other work only considers H$\alpha$
\emph{emission} as indicator for activity. In those cases where
\citet{Jenkins:2009p2650} detected H$\alpha$ in emission, we do not
provide any value in our table. The stars can easily be identified in
the catalogue as those stars from \citet{Jenkins:2009p2650} that have
no value of $\log{\rm H_{\alpha}}/L_{\rm bol}$ in our catalogue.

Normalized luminosities or upper limits of it are available for 244 stars
(73\,\% of our total sample of 334).

\section{Analysis}
\label{sec:analysis}

Normalized H$\alpha$ luminosity was calculated for our new spectra as
a proxy for chromospheric activity, and projected rotational velocity
was determined using the cross-correlation method.

\subsection{Chromospheric activity}
\label{ssec:halpha_det}

Chromospheric activity is measured from H$\alpha$ emission in our spectra. We
estimated the continuum around H$\alpha$ taking the median of two different
regions on either sides of H$\alpha$ line, namely: 6545--6559\,\AA\ and
6567--6580\,\AA. This value is used to normalize the spectra. We integrated the
equivalent width of H$\alpha$-emission in the spectral range 6552--6572\,\AA.

Many stars of our sample do not show significant line emission at
H$\alpha$. We conservatively estimated the detection limit of our
spectra to 0.2\,\AA, which is consistent with the approach of
\citet{Cayrel:1988p2705} assuming a 3\,$\sigma$ detection limit with
\begin{equation}
  \sigma_{\mathrm{EqW}} = 1.5 \frac{\sqrt{\mathrm{FWHM}_{line} \delta x_{line}}}{\mathrm{S}/\mathrm{N}}
\end{equation}
and a typical SNR of 50. In this equation, FWHM$_{line}$ is the full width
half maximum of the expected line and $\delta x_{line}$ is the size of a
resolution element. 

Equivalent width is not a suitable indicator of stellar activity because the
continuum flux is a steep function of effective temperature. We used PHOENIX
model atmospheres \citep[assuming $\log g = 5.0$][]{Hauschildt:1999p3036} to
transform equivalent widths to H$\alpha$ flux, $F_{\rm H_{\alpha}}$, and we
determine the flux ratio $F_{\rm H_{\alpha}}/F_{\rm bol}$ using $F_{\rm bol} =
\sigma T^4$. Effective temperatures were derived from spectral type using the
conversion given by \citet[]{Kenyon:1995p2462}. Finally, we use the identity
$F_{\rm H_{\alpha}}/F_{\rm bol} = L_{\rm H_{\alpha}}/L_{\rm bol}$ to determine
the ratio between H$\alpha$ and bolometric luminosity, i.e., normalized
H$\alpha$ luminosity.

\subsection{Rotation}
\label{ssec:vsini_det}

\begin{figure}
  \plotone{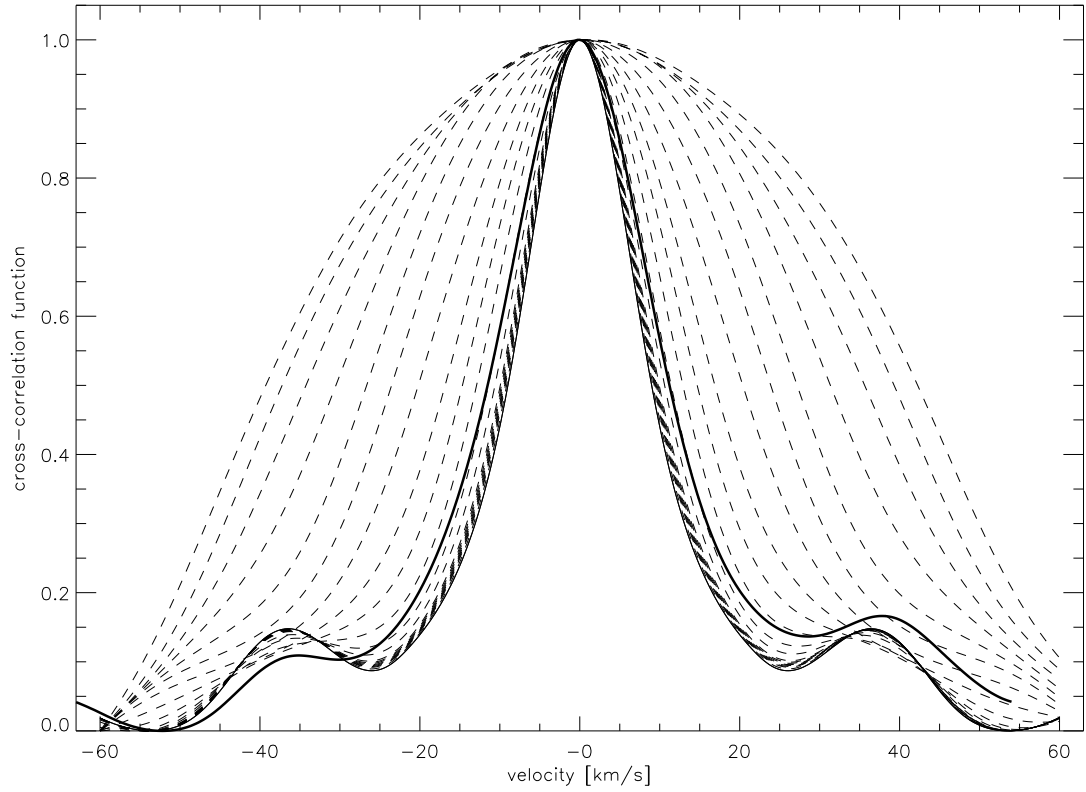}
  \caption{Typical cross-correlation profiles. Dashed lines show
    cross-correlation functions of artificially broadened spectra with the
    template spectrum. The dark line is the cross-correlation function from an
    object spectrum with the template spectrum.}
  \label{plot:xcorr_fr}
\end{figure}

In order to derive projected rotational velocities, $v\,\sin{i}$, from
our spectra, we used the cross-correlation method as mentioned
above. As a first step, the spectrum of a slowly rotating star was
chosen as a template spectrum and artificially broadened according to
a set of different velocities \citep[]{2005oasp.book.....G}. We chose
velocities in the range $[1,40]$\,km\,s$^{-1}$, limb darkening was set
to $0.6$ \citep[see][]{Browning:2010p2615}. The cross-correlation
functions between the unbroadened template spectrum and the set of
broadened spectra was calculated and the FWHM were determined as a
function of $v\,\sin{i}$. To derive $v\,\sin{i}$ in a target spectrum,
the cross-correlation function between the object spectrum and the
template was calculated, an example is shown in
Fig.\,\ref{plot:xcorr_fr}. The FWHM is measured and converted into
$v\,\sin{i}$ according to the calibration established from the
broadened template spectra.

The two spectrographs used for our observations have somewhat
different spectral resolving power. In order to avoid systematic
differences between the two data sets, we employed different template
spectra, i.e., one for each instrument. For the FOCES sample, we used
a spectrum of Gl~2 as a template star, for the FEROS sample, Gl~84 was
used. Both objects are of spectral type M2. We tried to use a few
different template stars but found no systematic differences.

We estimated the detection limit for rotational broadening from our
procedure to determine $v\,\sin{i}$. If we artificially broadened the
spectra to simulate slow rotation ($v\,\sin{i} \la 3$\,km\,s$^{-1}$),
the FWHM of the cross-correlation profile only marginally differed
from the auto-correlation function of the template. In our case, the
spectral resolving power is not high enough to fully resolve the lines
of slow rotators so that the threshold at which significant broadening
becomes visible is determined by the spectral resolving power of the
instrument. We find that from the FOCES spectra ($R = 40,000$) we can
determine values of $v\,\sin{i}$ in excess of $4$\,km\,s$^{-1}$. For
the FEROS spectra ($R = 48,000$), the detection limit is at
$v\,\sin{i} = 3$\,km\,s$^{-1}$.

For the cross-correlation procedure, we use only selected spectral
regions that are virtually free of telluric absorption and emission
lines. We calculated correlation functions in 13 different spectral
regions, each covering approximately 20\,\AA. The spectral regions are
not identical in the FEROS and FOCES samples, which is due to the
different performances of the instruments and their coverage of
\'{e}chelle orders. The projected rotational velocity was calculated
for each spectral region; the adopted final rotational velocity is the
mean of the individual values. The standard deviation for stars with
$v\sin i < 20$\,km\,s$^{-1}$ is typically below
$1$\,km\,s$^{-1}$. Note that this value is much smaller than the
detection limit because it is the typical scatter in $v\,\sin{i}$
between individual orders while the detection limit is the lowest
value of $v\,\sin{i}$ at which rotation can be distinguished against
other broadening agents like temperature and instrumental
broadening. The uncertainty of our $v\,\sin{i}$ measurements also is
not the same as the intra-order scatter, it is the accuracy at which
we can distinguish rotation from other broadening agents; we estimate
the final uncertainty to be $\sim 3$\,km\,s$^{-1}$ for slow rotators
but not less than 10\,\% \citep[see][]{Reiners:2007p18,
  Reiners:2007p19}.

\subsection{Spectroscopic binaries}\label{ssec:spec_bin}

\begin{deluxetable}{l c c c c}
\tablewidth{0.0cm}
\tablecaption{Spectroscopic Binaries excluded from the Analysis\label{tab:sb}}
\tablehead{\colhead{Name} & \colhead{$\alpha (J2000)$} & \colhead{$\delta (J2000)$} & \colhead{Spectral Type} & \colhead{Reference}}
\startdata
Gl 1054A  & 03 07 56.0	 & $-$28 13 09	 &  M 0.0  & (1,2) \\ 
Gl 508A   & 13 19 45.5	 & +47 46 39	 &  M 0.5  & (1,2) \\ 
Gl 29.1   & 00 42 47.9   & +35 32 54     &  M 1.0  & (1,3) \\
Gl 616.2  & 16 17 05.4	 & +55 16 11	 &  M 1.0  & (1) \\
G 140-009 & 17 43 00.6   & +05 47 21     &  M 1.0  & (1) \\
Gl 804    & 20 44 21.9	 & +19 45 01	 &  M 1.0  & (1) \\ 
Steph 1806& 20 40 56.4	 & $-$10 06 47	 &  M 1.5  & (1) \\
Gl 815A   & 21 00 04.9	 & +40 04 14	 &  M 1.5  & (1,2) \\ 
Gl 54     & 01 10 21.2	 & $-$67 26 54	 &  M 2.5  & (1) \\
Gl 268.3  & 07 16 19.7	 & +27 08 33	 &  M 2.5  & (1) \\
Gl 644A   & 16 55 28.1	 & $-$08 20 16	 &  M 3.0  & (1,4) \\ 
Gl 735    & 18 55 27.3	 & +08 24 09	 &  M 3.0  & (1) \\ 
Gl 206    & 05 32 14.6	 & +09 49 15	 &  M 3.5  & (1) \\
G 097-052 & 05 34 15.1	 & +10 19 15	 &  M 3.5  & (1) \\ 
Gl 263    & 07 04 17.2	 & $-$10 30 08	 &  M 3.5  & (1) \\
GJ 2069A  & 08 31 37.6	 & +19 23 39	 &  M 3.5  & (1,4) \\ 
Gl 375    & 09 58 33.3	 & $-$46 25 23	 &  M 3.5  & (1) \\
GJ 1212   & 17 13 40.6	 & $-$08 25 11	 &  M 3.5  & (1) \\
LP 476-207& 05 01 58.7	 & +09 58 59	 &  M 4.0  & (1,4) \\
LHS 2887  & 14 17 03.2	 & +31 42 47	 &  M 4.0  & (1,4) \\ 
\tableline
Gl 381  & 10 12 04.3	 & $-$02 41 00	 &  M 2.5  & (4)  \\
GJ 1080 & 05 28 14.8	 & +02 58 23	 &  M 3.0  & (5)  \\
Gl 487  & 12 49 03.1	 & +66 06 37	 &  M 3.0  & (4)  \\
Gl 747(A) & 19 07 42.1	 & +32 32 32	 &  M 3.0  & (4)  \\
LHS 6158 & 08 58 56.1	 & +08 28 28	 &  M 3.5  & (4)  \\ 
G 203-47  & 17 09 31.2	 & +43 40 54	 &  M 3.5  & (4) \\
Gl 661(A) & 17 12 07.5	 & +45 40 09	 &  M 3.5  & (4)  \\
Gl 896A \tablenotemark{a} & 23 31 51.8	 & +19 56 14 &  M 3.5 & (4) \\ 
Gl 695BC & 17 46 27.2	 & +27 43 07	 &  M 3.5 & (4)  \\ 
GJ 3129 & 02 02 44.0	 & +13 34 33	 &  M 4.5 & (5)  \\
Gl 268 & 07 10 07.8 & +38 31 27	 &  M 4.5  & (2) \\
GJ 1103(A) & 07 51 56.7	 & $-$00 00 08	 &  M 4.5  & (4)  \\
LHS 3080 & 15 31 54.4	 & +28 51 08	 &  M 4.5 & (5)  \\
GJ 1230A & 18 41 09.3	 & +24 47 14	 &  M 4.5 & (4)  \\ 
Gl 831(A) & 21 31 17.8	 & $-$09 47 25	 &  M 4.5 & (4)  \\
\enddata
\tablenotetext{a}{Gl 896B is located less than 1\AA\ from the A component. It has measured $v\sin i$ of $24.2$ \,km\,s$^{-1}$ in \cite{Mohanty:2003p83}. It cannot be confirmed whether the estimates are for both components combined.}
\tablerefs{
(1) New Observations; (2) \citet{2002AJ....123.3356G}; (3) \cite{Browning:2010p2615}; (4) \cite{Delfosse:1998p79}; (5) \cite{Jenkins:2009p2650}.}
\end{deluxetable}

\begin{figure}
  \plotone{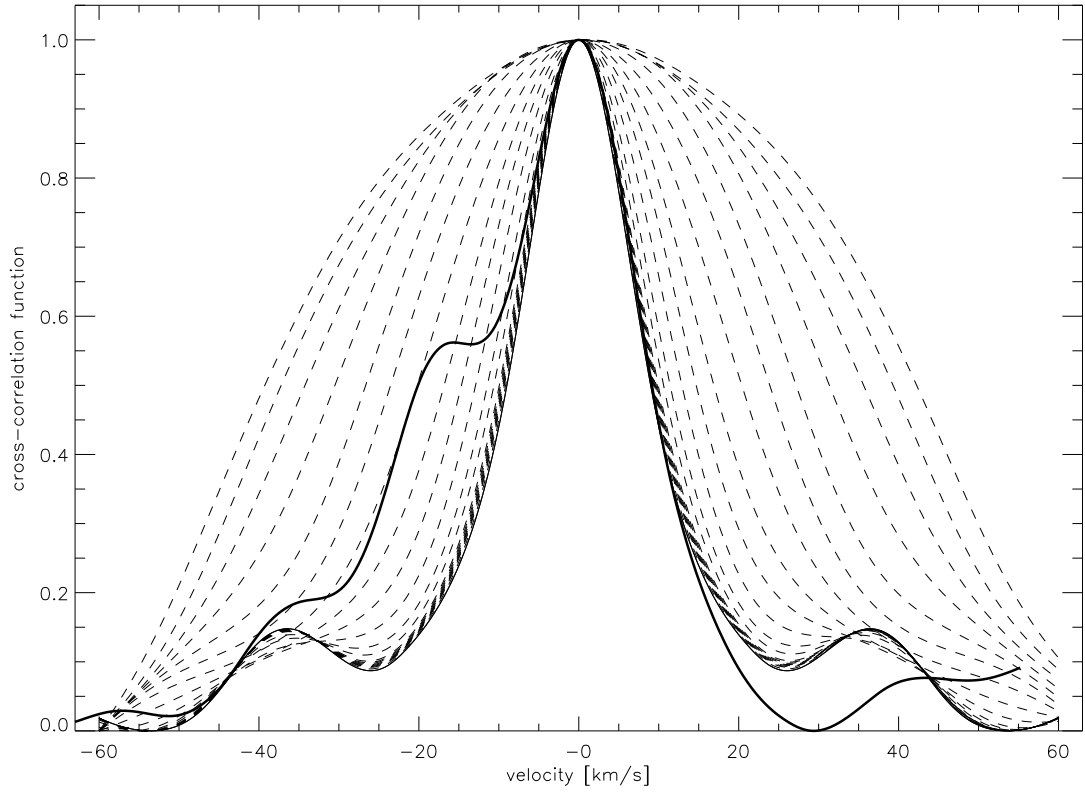}
  \caption{\label{plot:xcorr_sb}Cross-correlation profile of a probable
    spectroscopic binary. Dashed lines show cross-correlations function of
    artificially broadened spectra with the template spectrum. The dark line
    is the cross-correlation function from an object spectrum with the
    template spectrum.}
\end{figure}

Complete information on binarity in stars is difficult to
achieve. Binarity may be detected using high spatial resolution
imaging, but in many cases binary components are too close to each
other and cannot be spatially resolved. In such a case, the observed
spectrum consists of light from all components weighted according to
their luminosity. If both components are similar in luminosity, both
spectra appear in the spectrum with a separation according to the
difference in radial velocities at the time of observation. For the
search for spectroscopic binaries, the cross-correlation profile is
very useful. Three cases can be distinguished: 1) The separation is
larger than the typical line-width. In such a case, two systems of
spectral lines are visible in the spectrum and the cross-correlation
profile shows two separated maxima; 2) The separation is on the order
of the typical line-width. Here, the cross-correlation function is
broader than individual correlation maxima due to single stars. The
profile may appear asymmetric depending on the luminosity difference
of the components and their radial velocity
difference. Fig.\,\ref{plot:xcorr_sb} shows a typical
cross-correlation profile with significant asymmetry that is
attributed to a spectroscopic binary; 3) The separation is marginally
different from zero. The profile appears only slightly wider than the
typical single profile and may be asymmetric.

If a system is a multiple system instead of a single star, the determination
of rotation becomes meaningless unless individual components can be
disentangled from each other. We found that 20 stars of our originally
observed sample are spectroscopic binaries that could be unambiguously
identified. The stars are listed in Table\,\ref{tab:sb} and are not used for
further analysis and the catalogue.  For completeness, Table\,\ref{tab:sb}
also includes spectroscopic binaries of spectral types M0.0--M4.5 reported in
the literature.

\begin{deluxetable}{l c c c}
  \tablewidth{0.0cm}
  \tablecaption{Possible Spectroscopic Binaries Excluded from the Analysis\label{tab:marginal_outliers}}
  \tablehead{\colhead{Name} & \colhead{$\alpha (J2000)$} & \colhead{$\delta (J2000)$} & \colhead{Spectral Type}}
  \startdata
  G 235-020 & 09 19 23.0	 & +62 03 18	 & M 0.0 \\
  Gl 341 & 09 21 40.2	 & $-$60 16 58	 & M 0.0 \\
  Gl 373 & 09 56 08.9	 & +62 47 21	 & M 0.0 \\
  GJ 1181A & 13 55 02.8	 & $-$29 05 23	 & M 0.0 \\
  Gl 737A & 18 57 30.6	 & $-$55 59 19	 & M 0.0 \\
  LTT 8848 & 22 05 51.3	 & $-$11 54 48	 & M 0.0 \\
  GJ 1135 & 10 41 08.9	 & $-$36 53 42	 & M 0.5 \\
  BPM 11774 & 18 50 25.7	 & $-$62 03 02	 & M 0.5 \\
  G 144-016 & 20 37 20.8	 & +21 56 54	 & M 0.5 \\ 
  Gl 207.1 & 05 33 44.8	 & +01 56 43	 & M 2.5 \\
  GJ 1136A & 10 41 50.6	 & $-$36 37 53	 & M 2.5 \\ 
  LHS 221A & 06 54 03.7	 & +60 52 24	 & M 3.0 \\
  Gl 352A & 09 31 18.9	 & $-$13 29 18	 & M 3.0 \\
  \enddata
\end{deluxetable}

Our cross-correlation analysis revealed that 13 other stars have asymmetric
cross-correlation profiles. The degree of asymmetry is small but justifies the
assumption that the stars are no regular single objects. They may be
spectroscopic binaries with long periods or observed at very similar radial
velocities. We exclude these objects from our sample analysis. The marginal
outliers are listed in Table\,\ref{tab:marginal_outliers}, further
observations at a different epoch may clarify whether these stars are in fact
binaries.

\section{Results}\label{sec:results}

Our catalogue of early- to mid-M dwarfs with measured rotational velocities
together with information about activity is given in
Table\,\ref{table:objects}. Spectral types are taken from
\citet{Reid:1995p1938}. Information on activity and rotation from our
observations and the literature are given as explained in the foregoing
sections.

\subsection{Chromospheric activity}\label{ssec:chromo_cs}

\begin{figure}
  \plotone{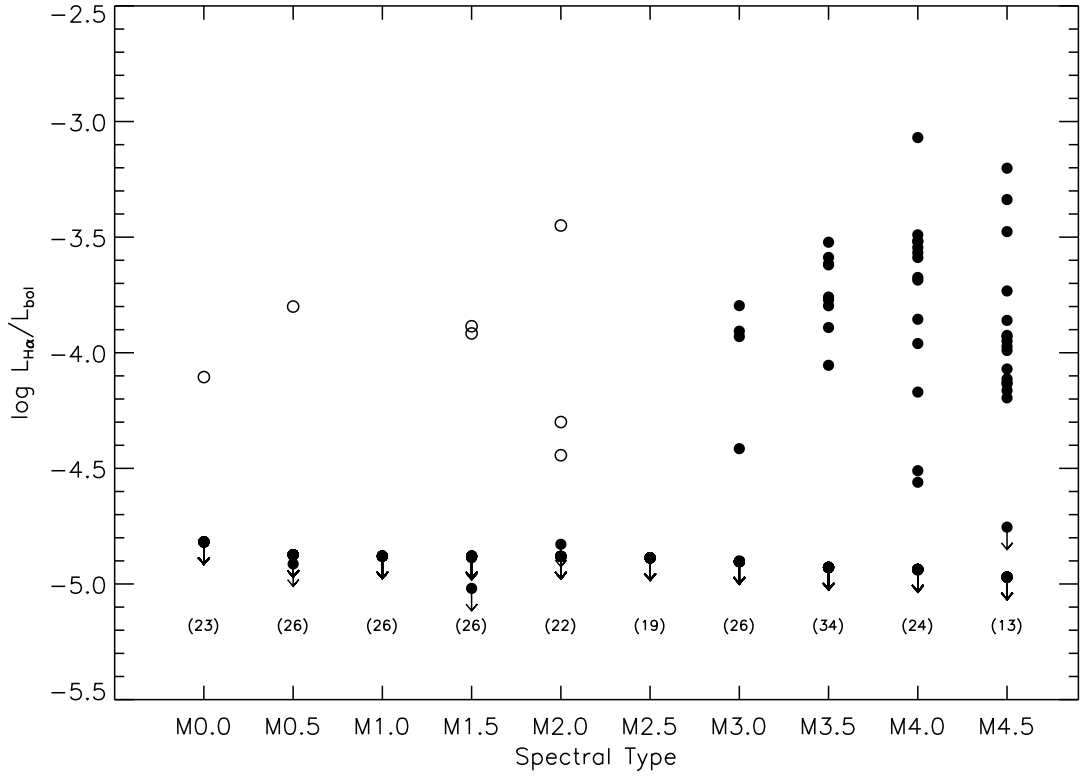}
  \caption{\label{plot:halpha_utsample}Normalized H$\alpha$ luminosities as a
    function of spectral type. Early-M dwarfs ($<$M3) with significant
    H$\alpha$ detections are shown as open circles, all other targets as full
    circles. Non-detections of H$\alpha$ are plotted at their detection levels
    with downward arrows added at their position. Numbers in parentheses show
    the number of non-detections per spectral bin that are often overplotted
    at the same position.}
\end{figure}

In Fig.\,\ref{plot:halpha_utsample}, we show normalized H$\alpha$
activity as a function of spectral type. In total, our catalogue
contains 244 stars with H$\alpha$ measurements, 95 of them (39\,\%)
are active. It is well established that activity lifetimes are
substantially longer at later spectral types
\citep[e.g.,][]{Hawley:1996p1924, 2002AJ....123.3356G,
  2005AJ....129.2428S, 2008AJ....135..785W}. Stars on the cool side of
the boundary to complete convection appear active much longer, leading
to the observation that many more fully convective stars show
activity. On the hot side of that boundary, where stars are believed
to still harbor a tachocline and hence may drive a Sun-like
large-scale dynamo, only very few active stars are known in the
field. Virtually all active early-M stars are members of young
associations, several examples can be found in
\citet{2006A&A...460..695T}. Partially convective early-M stars in the
field, however, that are believed to be older, in general do not
possess significant activity. Within the literature concerned for our
catalogue, there is no early-M type star ($<$M3) with significant
activity that is not a known member of a young association. Thus, any
single early-M type active star is probably young, which means not
older than a few 100\,Myr. In fully convective stars, however,
activity can persist for several Gyr and we expect to find many more
active stars of spectral type M3 and later. Note that rapid rotation
and hence enhanced activity may be maintained in tidally locked
binaries.

\begin{figure}
  \plotone{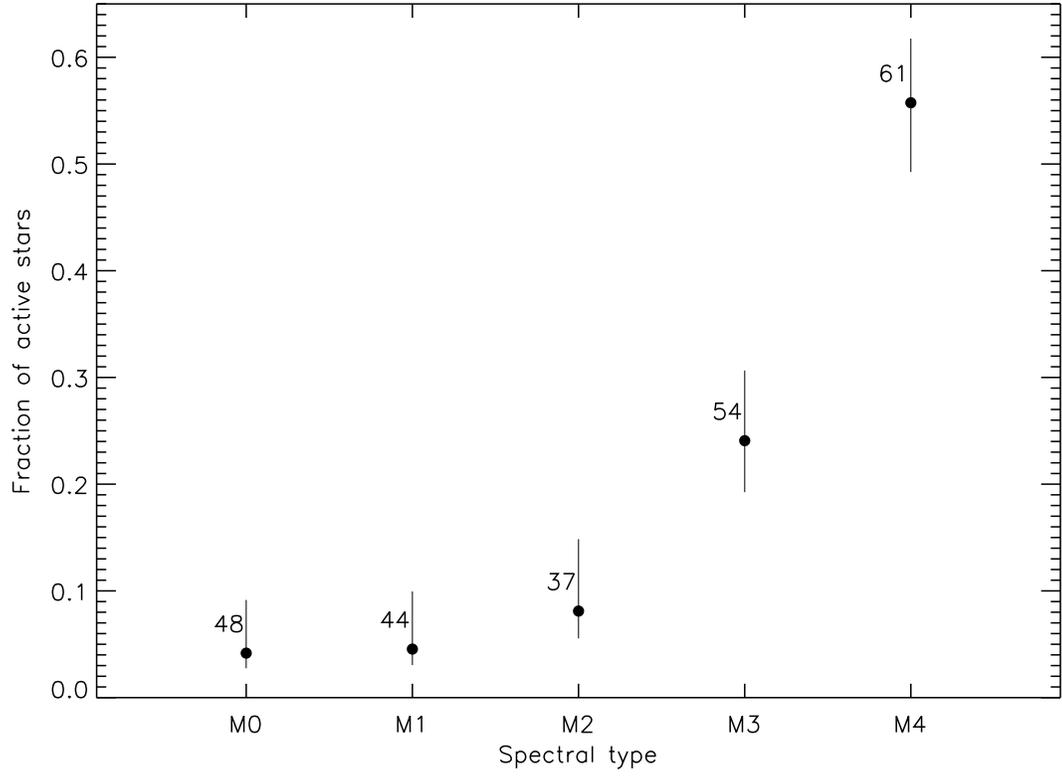}
  \caption{\label{plot:actfrac_utsample}Fraction of active stars per spectral
    type in our sample. Numbers show how many stars are measured per spectral
    bin. Error bars show 1$\sigma$-uncertainties.}
\end{figure}

The total catalogue contains seven out of 129 (5\,\%) early-M type
stars exhibiting significant activity as H$\alpha$ emission. In
contrast, 47 out of 115 (41\,\%) stars between M3 and M4.5 are
active. The fraction of active stars as function of spectral type is
shown in Fig.\,\ref{plot:actfrac_utsample}, it is in general higher
than reported in \citet{2008AJ....135..785W} from the SDSS sample,
which is consistent with our sample being younger than the sample used
there (observed away from the Galactic plane), and our results are
consistent with earlier work on the activity fraction of field stars
\citep[e.g.,][]{Hawley:1996p1924, 2002AJ....123.3356G,
  2005AJ....129.2428S}. The general trend, however, is very well
reproduced. A relatively sharp transition from a low activity fraction
smaller than 10\,\% to a significant fraction above 50\,\% occurs
around spectral type M3. From activity information alone, we cannot
determine the reason for this dramatic
increase. \citet{2008AJ....135..785W} speculate that longer activity
lifetimes may be explained by a transition from a
rotationally-dependent solar-like dynamo to a rotationally-independent
turbulent dynamo in which magnetic fields can survive much longer even
if the stars are rotating slower. We come back to this point when we
discuss the rotation of these stars in Sect.\,\ref{ssec:rot_cs}.

\subsection{Active early-M dwarfs}
\label{sect:active_early}

Field stars are believed to be relatively old ($\sim$Gyr) and early-M dwarfs
($<$M3) are in general not observed to be active in the field. In our survey,
we found or confirmed activity in seven out of 129 (5\,\%) early-M targets
with H$\alpha$ measurements. The reason for their activity is probably youth
since it is known that early-M dwarfs can be very active at young ages. Young
stars are rotating much more rapidly and therefore can generate sufficient
magnetism to generate magnetic activiy \citep{Pizzolato:2003p74}. A potential
reason why a field early-M dwarf can generate activity is prevention of
angular momentum loss because of binarity. Another explanation is that the
star is indeed young and entered our survery because it is relatively nearby
compared to other young objects. Finally, some stars may be mis-classified in
spectral type so that they are actually within the regime of M3 or later.

The seven active early-M dwarfs found in our survey are presented in
Table\,\ref{table:active_early}. Spectral types are between M0 and M2,
mis-classification may be an issue for the latest targets but is
unlikely for all of them. Some of the stars have companions but we did
not find evidence for binarity in any of the stars that would
sufficiently influence the rotation of the star on
Gyr-timescales. Most of the objects are probably young objects in the
solar neighborhood. We discuss the seven stars individually in the
following:

\begin{deluxetable}{lccccr}
  \tablewidth{0pt}
  \tablecaption{\label{table:active_early}Active early-M stars}
  \tablehead{Name & Spectral type & \multicolumn{2}{c}{log($L_{\rm
        H_{\alpha}}/L_{\rm bol}$)} & \multicolumn{2}{c}{$v\sin i$}\\
  &&&&\multicolumn{2}{c}{[km/s$^{-1}$]}}
  \startdata
  \input{EarlyActive_table.tex}
  \enddata
\end{deluxetable}

\medskip

\noindent \textbf{Gl 182} is a young star known as V1005~Ori. The star is
contained in the SACY sample \citep{2006A&A...460..695T, daSilva:2009p2685}
and classified as a member of the $\beta$ Pictoris young association
(10\,Myr). It shows substantial activity ($\log{L_{\rm H_{\alpha}}/L_{\rm
    bol}} = -4.11$) and rapid rotation ($v\,\sin{i} = 10.4\,$km\,s$^{-1}$).
	
\medskip

\noindent \textbf{Gl 494A} has a companion of spectral type M7 at a
separation of $0.475 ''$ \citep[]{Beuzit:2004p2678}, and a planetary
candidate companion of spectral type T8--9 at $102 ''$
\citep{Goldman:2010p3162}. The secondary component is too faint to
influence the rotational profile and our measurement of $v\,\sin{i}$
is most likely the one of the M0.5 primary
alone. \citet[]{Beuzit:2004p2678} determine $v\,\sin{i} =
9.6$\,km\,s$^{-1}$ consistent with our observations. They conclude
that the object is not a short-period locked binary but a rapidly
rotating, young early-M star \citep[see also][]{2010ApJ...725.1405B,
  2011MNRAS.414.3590B}.

\medskip

\noindent \textbf{Steph 546A} (GJ\,3331A) also is contained in the SACY survey
under the name BD~21\,1074A. It is classified as a member of the $\beta$~Pic
association \citep{daSilva:2009p2685} and has a companion pair,
BD~21\,1074BC.

\medskip

\noindent \textbf{Wo 9520} (GJ\,9520) was observed by
\citet[]{Daemgen:2007p2687} using the Altair AO System at Gemini North
Observatory. No companion could be detected. \citet[]{Shkolnik:2009p2686}
observed the star at two different epochs and found no evidence for RV
variability providing evidence that Wo\,9520 is not a binary
system. \citet[]{Shkolnik:2009p2686} estimate an age of $15-150$ Myrs.

\medskip

\noindent \textbf{GJ 2036A} (CD-56 1032A) is part of a binary system with two
active components. The system is classified as a member of the AB~Doradus
young association ($70$ Myrs) by \citet{daSilva:2009p2685}.

\medskip 

\noindent \textbf{Gl 358} is identified as a possible member of the
Carina-Near Stream by \citet{2006ApJ...649L.115Z}, which would imply an age of
$\sim 200$\,Myr.

\medskip 

\noindent \textbf{Gl 569A} is accompanied by the brown-dwarf brown-dwarf pair
Gl\,569Bab \citep{2006ApJ...644.1183S, 2011MNRAS.413.1524F} for that orbital
parameters are well determined. Comparison of the colors of Gl\,569Bab to
theoretical isochrones and color mass diagrams suggest an age of
100--125\,Myr, which is probably the same as for Gl\,569A. The orbital
inclination of Gl\,569Bab is $(32.4 \pm 1.3)^{\circ}$ \citep[$\sin{i} = 0.54$;
][]{2006ApJ...644.1183S}.

\medskip

Five out of the seven active early-M stars show detectable rotation in
$v\,\sin{i}$. The two other stars, in which no rotation was detected, Gl~358
and Gl\,569A, may be observed under high inclination. Both stars show
relatively low activity compared to the other stars in
Table\,\ref{table:active_early} so that their equatorial velocities are
probably comparably low, i.e. only a few km\,s$^{-1}$. Inclination angles
below $\sim 60^{\circ}$ may be sufficient to push $v\,\sin{i}$ below the
detection limit, which renders this scenario rather likely. In particular,
this is consistent with the assumption of spin-orbit alignment in the Gl\,569
multiple system ($i \approx 30^{\circ}$).

For each of the seven active early-M stars, \citet{Kiraga:2007p407} provide a
photometric period. This may be a hint that spot configurations in active
early-M dwarfs are rather stable at least after a few 10\,Myr. However,
periodicity may not in all cases be due to rotation. We discuss photometric
periods in Sect.\,\ref{sect:periods}.



\subsection{Rotation}
\label{ssec:rot_cs}

We provide measurements or upper limits of the projected rotational velocity,
$v\,\sin{i}$, for all 334 stars of our catalogue. As explained above,
detection thresholds for our new measurements with the FOCES and FEROS
spectrographs are estimated at 4 and 3\,km\,s$^{-1}$, respectively, according
to the spectrographs' different resolving power. Detection limits of the
collected data from the literature also vary. \citet{Browning:2010p2615}
estimate a detection limit of $v\,\sin{i}_{\rm lim} = 2.5$\,km\,s$^{-1}$ ($R
= 45,000$--$60,000$), \citet{Reiners:2007p19} use $v\,\sin{i}_{\rm lim} =
3$\,km\,s$^{-1}$ ($R = 31,000$), and \citet{Reiners:2007p18} used extremely
high-resolution data ($R \approx 200,000$) estimating $v\,\sin{i}_{\rm lim} =
1$\,km\,s$^{-1}$. Furthermore, we adopt $v\,\sin{i}_{\rm lim} =
3$\,km\,s$^{-1}$ for results from \citet[][$R = 40,000$]{Marcy:1992p2101} and
$v\,\sin{i}_{\rm lim} = 2$\,km\,s$^{-1}$ for those from \citet[][$R =
42,000$]{Delfosse:1998p79} as written in those publications.
\citet{Jenkins:2009p2650} do not quote a general detection threshold in
$v\,\sin{i}$. Using spectra at a spectral resolving power of $R = 37,000$, they
determine individual upper limits as well as many detections at the
$v\,\sin{i} = 3$\,km\,s$^{-1}$ level. We adopt these values as they are given
in the original literature but show below that their detection threshold is
likely higher, around $v\,\sin{i}_{\rm lim} = 4$\,km\,s$^{-1}$.

Several stars are contained in more than one of the considered
works. Individual measurements may differ because spectral appearance can
change with chromospheric variability or different authors used different
spectral lines in their analysis that are sensitive to chromospheric
activity. We provide a list of all stars with more than one $v\,\sin{i}$
measurement in Table\,\ref{tab:vsinicomp}. In general, all measurements from
different data are consistent within the uncertainties, the only outliers are
two measurements of $v\,\sin{i}$ in Gl~388 and Gl~873 by
\citet{Delfosse:1998p79}. For the inactive stars Gl~369, G\,244-047.01, and
GJ~1119, rotational broadening is reported in one paper but not detected in
one or more other works. These stars are probably rotating very slowly as we
discuss below. Gl~388 (AD~Leo), Gl~729 (V1216 Sgr), and Gl~873 (EV~Lac) are
very active stars for that different analyses report $v\,\sin{i}$ values close
to the detection limits. These stars are probably rotating on the few
km\,s$^{-1}$ level.

\begin{figure}
  \plotone{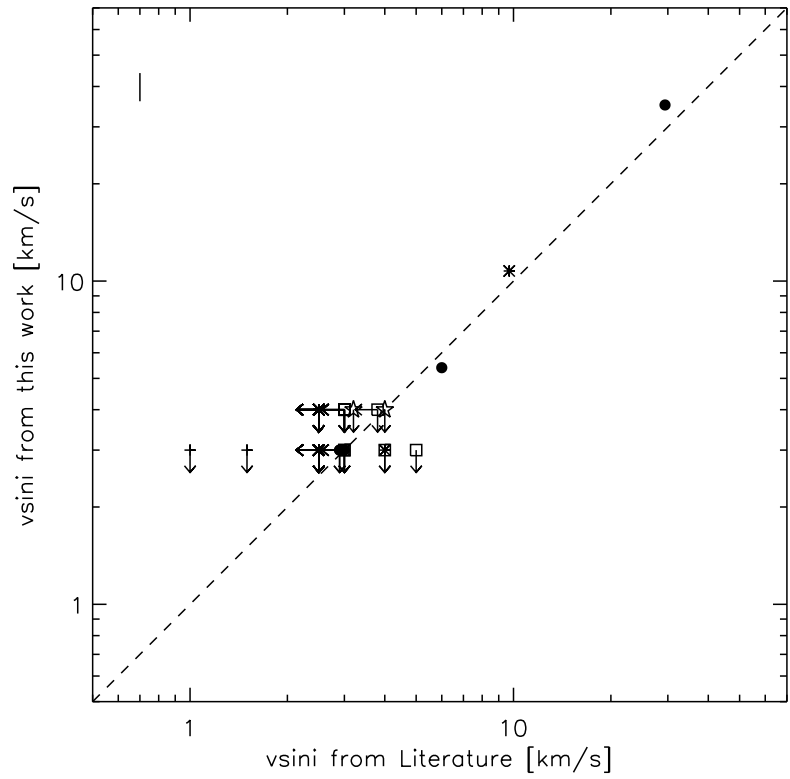}
  \caption{\label{fig:vsini_comp}Comparison between $v\,\sin{i}$
    values from this work and the literature. Different symbols are
    used for different literature sources (see
    Table\,\ref{tab:vsinicomp}).}
\end{figure}

For our catalogue, we adopted $v\,\sin{i}$ values according to the
following strategy. Preference was given to the $v\,\sin{i}$ data from
the work done with the highest spectral resolution. Highest priority
was given to results \citet{Reiners:2007p18} because they are derived
from the highest resolution spectra. Second highest priority is given
to the results from \citet{Browning:2010p2615}. If both are not
available and several other works provided measurements of
$v\,\sin{i}$, we chose to use the one from our new measurements. In
Fig.\,\ref{fig:vsini_comp}, we show a comparison between $v\,\sin{i}$
measured in this work and data from the literature. For all stars,
values are consistent within the uncertainties and detection limits.

\begin{figure}
  \plotone{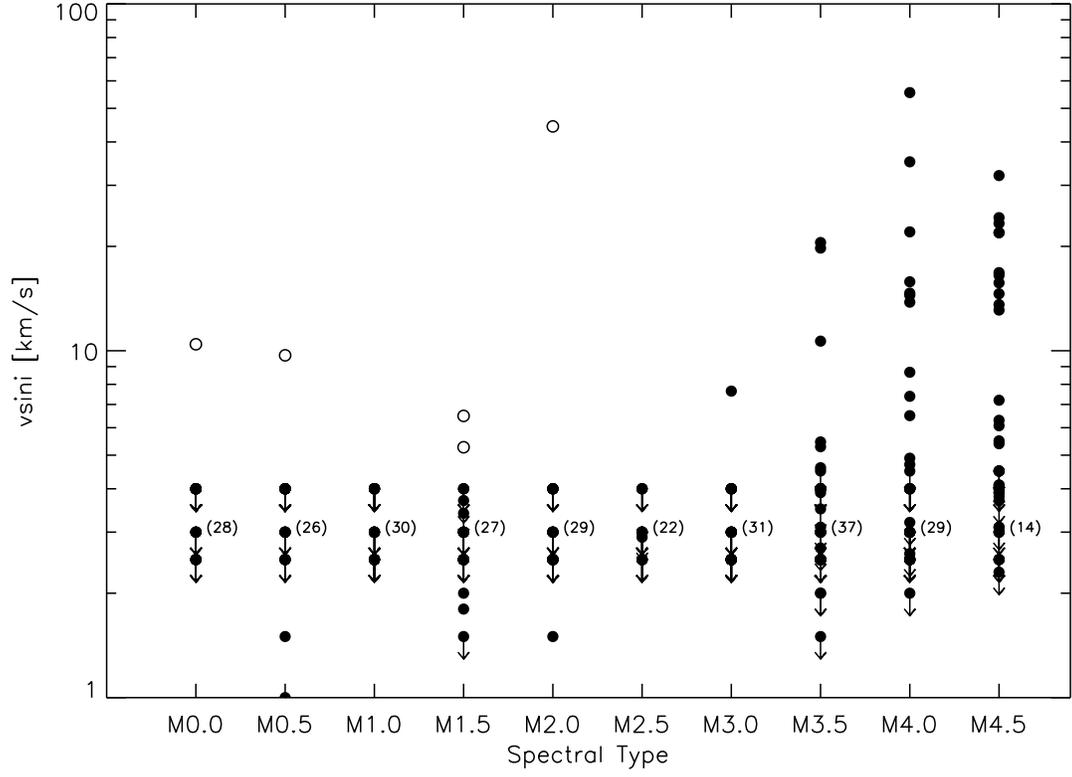}
  \caption{\label{plot:vsini_utsample}Projected rotational velocity ($v\sin
    i$) as a function of spectral type. Upper limits in $v\,\sin{i}$ are shown
    with downward arrows. Open circles show early-type M stars ($<$M3) that
    were found to be rotating faster than $v\,\sin{i} = 3$\,km\,s$^{-1}$ (five
    stars). The numbers in parentheses denote the numbers of slow rotators per
    spectral bin in which rotation is below the detection threshold (sum of
    all stars with downward arrows in this bin).}
\end{figure}

Projected rotational velocities, $v\,\sin{i}$, for our catalogue are plotted
as a function of spectral type in Fig.\,\ref{plot:vsini_utsample}. The
situation appears very similar to the one in Fig.\,\ref{plot:halpha_utsample}
where activity was shown as a function of spectral type. Again, we see only a
few early-M type stars with significant rotation (shown as open circles in
Fig.\,\ref{plot:vsini_utsample}). These stars are listed in
Table\,\ref{table:active_early} and discussed individually in
Sect.\,\ref{sect:active_early}. At later spectral types ($\ge$M3), significant
rotation appears to be more frequent just as activity is more frequent in this
spectral range. In total, 51 stars of our 334 stars (15\,\%) show rotational
broadening of $v\,\sin{i} \ge 3$\,km\,s$^{-1}$.

\begin{figure}
  \plotone{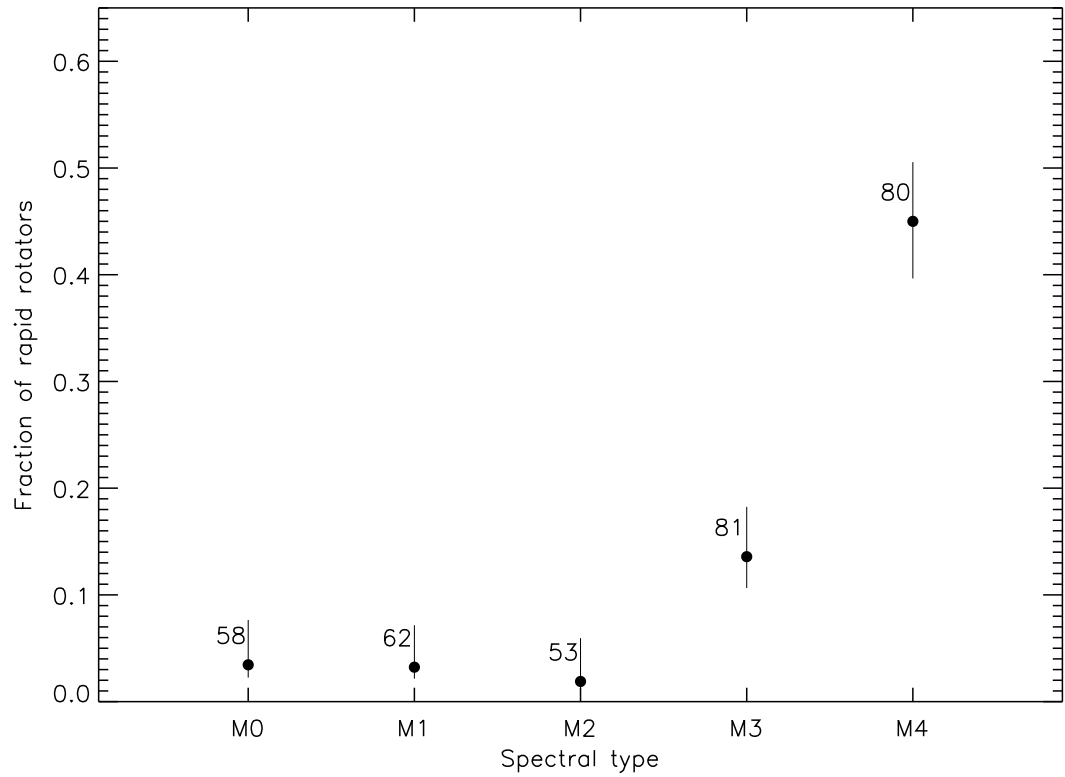}
  \caption{\label{plot:rotfrac_utsample}Fraction of rapid rotators per
    spectral type in our sample. Rapid rotators are stars with detected
    rotational broadening at $v\,\sin{i} = 3$\,km\,s$^{-1}$ or larger. Numbers
    show how many stars are measured per spectral bin. Error bars show
    1$\sigma$-uncertainties.}
\end{figure}

Fig.\,\ref{plot:rotfrac_utsample} shows the fraction of rapid rotators, i.e.,
stars with detected rotational broadening $v\,\sin{i} \ge 3$\,km\,s$^{-1}$, as
a function of spectral type. The picture appears to be very similar to the
fraction of active stars discussed above
(Fig.\,\ref{plot:actfrac_utsample}). While among early-M type stars ($<$M3),
the fraction of rapid rotators is below 5\,\%, it rises rapidly to
approximately 45\,\% at spectral type M4.

\subsection{Rotation-activity relation}
\label{sect:rotact}

After our discussions of activity and rotation in the sample
catalogue, we now turn to the relation between the two across the
spectral range M0.0--M4.5, i.e., from partially convective Sun-like
stars to fully convective stars. The two distributions in
Figs.\,\ref{plot:actfrac_utsample} and \ref{plot:rotfrac_utsample} are
strikingly similar. Within statistical uncertainties, the
distributions of active and rapidly rotating stars are consistent with
the assumption that active stars and rapid rotators are both drawn
from the same underlying population. We tested the probability that
both distribution are drawn from the same distribution following
Kolmogorov-Smirnov statistics \citep{1992nrca.book.....P}. In numbers,
the probability that the sample of active stars and the sample of
rapidly rotating stars are drawn from independent distributions is
lower than $4 \cdot 10^{-4}$. This means that rotation and activity
are highly correlated and that the active stars in our sample are
likely the same stars as the rotating ones. While this does not prove
a causal relation between rotation and activity, it provides evidence
that both occurs in the same stars.

It is worth emphazising that evidence for a correlation between
activity and rotation exists \emph{over the entire sample} including
both partially and fully convective stars.  In other words, before we
discuss the relation between rotation and activity on the basis of
individual stars, the distribution of rotation and activity in M0--M4
stars already provides strong evidence for the validity of the
rotation-activity connection across the boundary to complete
convection.

\begin{figure}
  \plotone{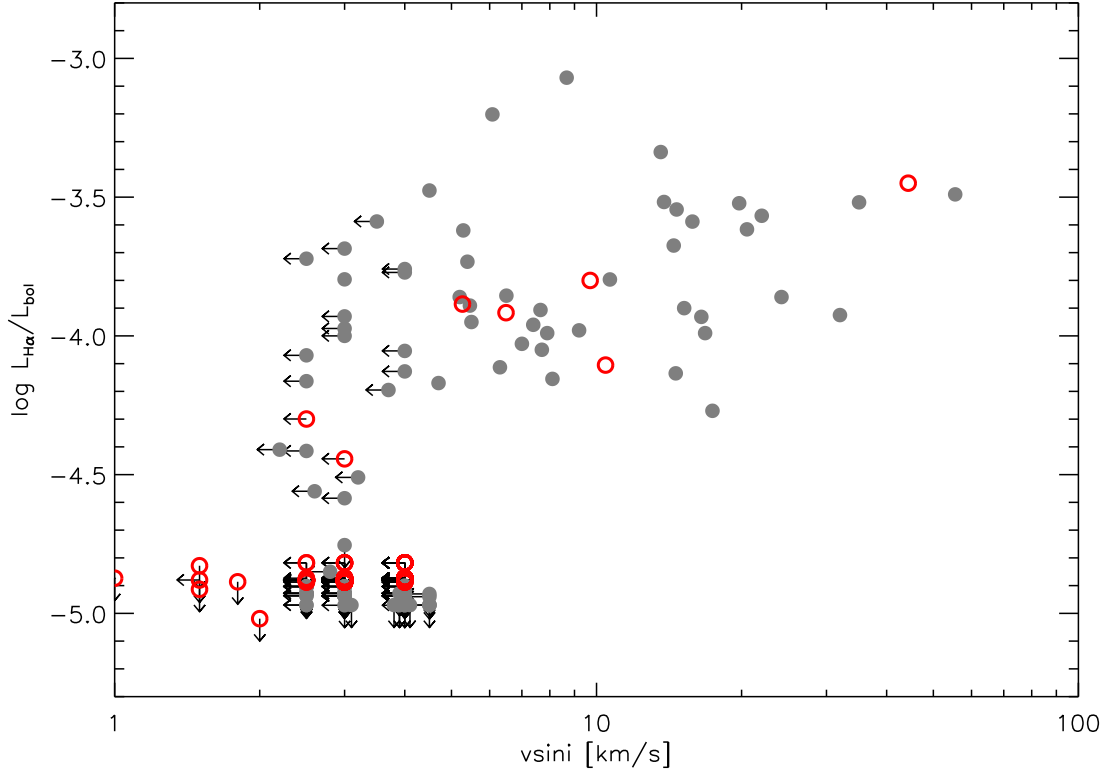}
  \caption{\label{plot:rotact_cs}Normalized H$\alpha$ luminosity as a function
    of projected rotational velocity ($v\sin i$). Open circles denote
    partially convective objects (M0.0--M2.5), whereas filled circles show
    fully convective ones (M3.0--M4.5). Leftward arrows show objects with
    upper limits in $v\,\sin{i}$, downward arrows show upper limits in
    H$_{\alpha}$ activity.}
\end{figure}

We plot normalized H$\alpha$ luminosity, $\log{L_{\rm H_{\alpha}}/L_{\rm
    bol}}$, against projected rotational velocity, $v\,\sin{i}$, in
Fig.\,\ref{plot:rotact_cs}. In the figure, we further discriminate between
partially convective stars ($<$M3) and likely fully convective stars (M3 and
later). The boundary between the two groups is likely not sharp and spectral
type uncertainties on the order of 0.5--1 spectral subtypes further softens
the location of this transition. As a first result, we can confirm the
rotation-activity relation in the sense that low activity ($\log{L_{\rm
    H_{\alpha}}/L_{\rm bol}} < -4.5$) only occurs at slow rotation. There are
a handful of inactive stars (stars with low activity) for that detections of
rotational line broadening on the order of 4--5\,km\,s$^{-1}$ is detected. We
discuss these stars in Sect.\,\ref{sec:rapid_and_inactive}. Furthermore, we
can conclude that the correlation between rotation and activity is valid at
both sides of the convection boundary, i.e., for both early- and later-M type
stars (open and solid symbols in Fig.\,\ref{plot:rotact_cs}).

Active stars are found at virtually all rotation rates. In our sample,
we found 48 very active stars with $\log{L_{\rm H_{\alpha}}/L_{\rm
    bol}} > -4.5$. Among them, 33 stars are rapid rotators
($v\,\sin{i} > 3$\,km\,s$^{-1}$). This means that 15 out of 48 active
stars (31\,\%) have rotation velocities below our detection limit, or
are observed under low inclination angles so that high rotation rates
are not detected. In the latter scenario, the most active stars at low
$v\,\sin{i}$ values are interpreted as stars observed under very low
angles $i$. We can test the assumption whether a tight relation
between rotation and activity is valid among all our sample stars. If
so, all stars with very high H$\alpha$ emission, say $\log{L_{\rm
    H_{\alpha}}/L_{\rm bol}} > -4.0$, would be rotating at
approximately $v\,\sin{i} \ga 5$\,km\,s$^{-1}$, which means that at a
typical detection threshold of $v\,\sin{i}_{\rm lim} =
3$\,km\,s$^{-1}$, inclination must be such that $\sin{i} < 0.6$ ($i <
37^{\circ}$). In our sample, 7 out of 36 (19\,\%) stars with
$\log{L_{\rm H_{\alpha}}/L_{\rm bol}} > -4.0$ show no detectable
rotation. In a sample of stars with randomly oriented rotation axes, a
fraction of 19\,\% will be observed at inclination angles smaller than
$36^\circ$, i.e., $\sin{i} < 0.59$. Thus, we can conclude that the
distribution of measurements in $v\,\sin{i}$ and activity is
consistent with the assumption of a well-defined relation between
rotation and activity that is spread out in Fig.\,\ref{plot:rotact_cs}
due to projection effects from observing the stars under statistically
distributed orientations. 

Active M stars are known to exhibit frequent flaring events that
introduce substantial scatter in $\log{L_{\rm H_{\alpha}}/L_{\rm
    bol}}$ and is difficult to quantify with just one
observation. \citet{2009AJ....138..633K} found that among a sample of
236 stars with flares, $\sim3\,\%$ show no H$\alpha$ emission outside
the flare. If we assume that inactive stars with occasional flaring
are slow rotators, we can estimate that one star of our 33 slowly
rotating flare stars is in fact an inactive, slow
rotator. Furthermore, \citet{2010AJ....140.1402H} determined the flare
rates of stars at different spectral types, for stars of our sample
the flare rates are $\le1\,\%$. Although it is not trivial to compare
the flare rates from \citet{2010AJ....140.1402H} to our sample because
of the inhomogenous distribution of exposure times, we can exclude a
significant influence of flaring on the occurence of active stars for
which no evidence of rotation could be detected.

In Fig.\,\ref{plot:rotact_cs}, we have distinguished between early-M
type stars ($<$M3, open symbols) and later M stars (solid symbols)
expecting the transition from partially to fully convective stars at
this spectral range. The subsample of early-M dwarfs is defined by a
number of inactive, slowly rotating stars (with several upper limits
in $v\,\sin{i}$) and seven active stars, five of them with detected
rotation. The early-M dwarfs exhibit a clear correlation between
projected rotation rate and normalized H$\alpha$ activity; all stars
with $v\,\sin{i} \approx 3$\,km\,$^{-1}$ and above are active, and the
stars with $v\,\sin{i} \ge 5$\,km\,$^{-1}$ show higher activity than
those at $v\,\sin{i} \approx 3$\,km\,$^{-1}$. Interstingly, the
scatter in activity in early-M stars at $v\,\sin{i} \ge
5$\,km\,$^{-1}$ is much smaller than the scatter in mid-M stars in our
sample. From our small sample, it is not possible to decide whether
this is an intrinsic effect or due to the small number of rapidly
rotating early-M stars, but it is consistent with the conclusion of
\citet{2002AJ....123.3356G} and \citet{2010ApJ...708.1482L} that the
H$\alpha$ variability is larger at later spectral types. In fully
convective star, the scatter of H$\alpha$ activity is much larger than
in early-M dwarfs, but as in early-M dwarfs all significantly rotating
fully convective show significant H$\alpha$ activity. In summary, the
saturation-type rotation-activity relation is intact until spectral
type M4.5, while the scatter is perhaps growing larger in lower-mass
stars that generally tend to be more active in our sample.

\subsection{Rapidly rotating inactive stars}
\label{sec:rapid_and_inactive}

\begin{deluxetable}{lcrlrc}
  \tablewidth{0pt}
  \tablecaption{\label{table:rapid_inactive}Inactive stars reported to
    be rapidly rotating. Stars marked with an asterisk show H$\alpha$ in
    absorption \citep{Jenkins:2009p2650}.}
  \tablehead{Name & Spectral type & \multicolumn{2}{c}{log($L_{\rm
        H_{\alpha}}/L_{\rm bol}$)} & \multicolumn{2}{c}{$v\sin i$}\\
  &&&&\multicolumn{2}{c}{[km\,s$^{-1}$]}}
  \startdata
  \input{FastInactive_table.tex}
  \enddata
\tablerefs{\footnotesize All $v\,\sin{i}$ values from \citet{Jenkins:2009p2650}}
\end{deluxetable}

The relation between rotation and activity may not be valid in all individual
cases. In general, rotation generates magnetic activity, but we may still find
activity in some stars that are observed at low \emph{projected} rotation
rates, or even in stars that are slowly rotating but active for reasons we
have not understood. On the other hand, our assumption of magnetic dynamo
operation triggered by rotation leads to the expectation that \emph{all}
rapidly rotating stars show significant values of magnetic
activity. \citet{2009ApJ...693.1283W} reported the existence of three rapidly
rotating but inactive stars at spectral types later than our sample
(M6--M7). We searched for such stars in our sample of hotter M stars.

Table\,\ref{table:rapid_inactive} lists nine stars in which no H$\alpha$
emission was found but a detection of rotational broadening was reported. All
nine stars are from the catalogue of \citet{Jenkins:2009p2650}, and all stars
have values of $v\,\sin{i}$ between 3 and 4.5\,km\,s$^{-1}$. Four of the nine
stars are reported to show H$\alpha$ in absorption, which is evidence for very
low but non-zero activity.

Regardless whether or not the stars with H$\alpha$ absorption are indeed
weakly active, it is striking that all nine stars are found in the same
work. Our catalogue contains 23 stars taken from \citet{Jenkins:2009p2650},
nine of them show no activity in the presence of rotation. Among the 311 stars
taken from other sources, not a single star shows similar properties. The
spectral resolution of the data used by \citet{Jenkins:2009p2650} is $R =
37,000$. For such data, a detection limit between 3 and 5\,km\,s$^{-1}$ can be
expected depending on SNR, that means at the same order as the measurements
reported for the stars in Table\,\ref{table:rapid_inactive}. We argue that the
absence of rapidly rotating inactive stars in the rest of our sample
provides ample evidence that \emph{all} rapid rotators $v\,\sin{i} \ga
3$\,km\,s$^{-1}$ show measurable H$\alpha$ activity, and that the reported
detections of rotation in inactive stars from \citet{Jenkins:2009p2650} are
spurious and in fact upper limits to their real values of $v\,\sin{i}$. Our
conclusion is that substantial chromospheric emission is a fundamental
consequence of rapid rotation in M0--M4.5 stars.

\section{Comparison to Photometric Periods}
\label{sect:periods}

Stellar line broadening provides information about the projected rotation
velocity on the surface of a star. A more convenient and physically meaningful
property is the rotation period of the star. Period $P$ and projected surface
velocity $v\,\sin{i}$ are related through

\begin{equation}
  v \sin{i} = \frac{2 \pi R \sin{i}}{P},
\end{equation}

with $R$ the stellar radius. Measuring photometric periods in M dwarfs
is notoriously difficult because of the high activy these stars reach
even at relatively long periods. This means that spot lifetimes may be
shorter than typical rotation periods. We collected photometric
periods from \citet{2011ApJ...727...56I} and \citet{Kiraga:2007p407},
the latter including reference to period measurements for Gl~411
\citep{Noyes:1984p48} and Gl~699
\citep{1998AJ....116..429B}. Furthermore, we consider periods
collected in \citet{2003A&A...410..671M}, namely periods for Gl~410
\citep{2000AJ....120.3265F} and Gl~735 \citep{1998ARep...42..649A},
and periods presented in \citet{2009AIPC.1135..221E} including the
period of Gl~873 from \citet{1995A&A...300..819C}. We augment our
sample of M0.0--M4.5 stars with three additional stars of spectral
type M5, GJ~1057 and GJ~1156, for which rotational periods are
reported by \citet{2011ApJ...727...56I}, and Gl~551 \citep[period
from][]{Kiraga:2007p407}, and in which measurements of $v\,\sin{i}$
are available. From the period, we estimated the star's surface
velocity for which information on stellar radius is
required. \citet{2011ApJ...727...56I} provide radius estimates for
their targets. For the stars in \citet[][including Gl~411 and
Gl~699]{Kiraga:2007p407}, we adopted the stellar masses provided there
and assumed that stellar radius in solar units has the same value as
stellar mass expressed in solar units \citep[i.e., for a star with $M
= 0.5$\,M$_{\sun}$ we assumed $R = 0.5$\,R$_{\sun}$;
see][]{2009A&A...505..205D}. For the other stars we used the strategy
of \citet{Kiraga:2007p407}, calculating mass from the $V$-band
mass-luminosity relation in \citet{2000A&A...364..217D} and assumed
mass-radius identity as above.

\begin{figure}
  \plotone{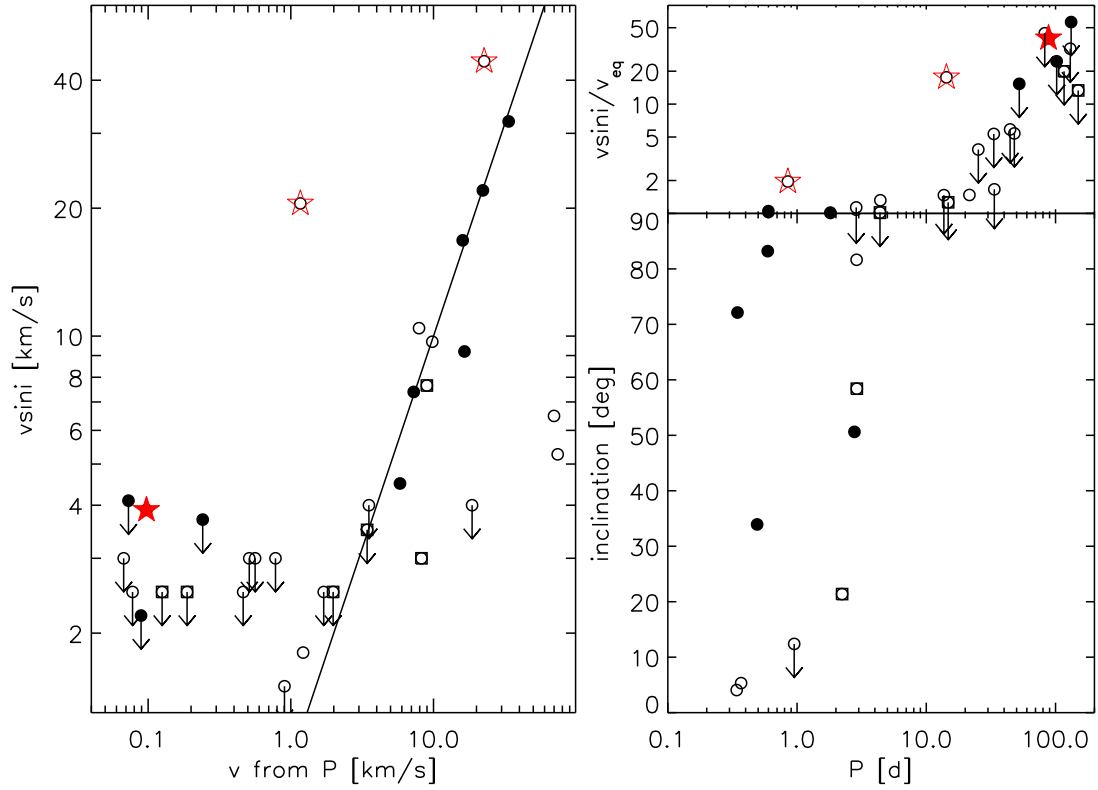}
  \caption{\label{fig:periods}\emph{Left panel:} Projected rotational
    velocity $v\,\sin{i}$ against surface equatorial velocity $v_{\rm
      eq}$ calculated from photometric period. \emph{Right panel:}
    Inclination angle derived from the comparison between $v\,\sin{i}$
    and $v_{\rm eq}$ if $v\,\sin{i} < v_{\rm eq}$ (lower panel), and
    ratio $v\,\sin{i} / v_{\rm eq}$ if $v\,\sin{i} > v_{\rm eq}$
    (upper panel). Downward arrows indicate upper limits in all values
    that are due to upper limits in $v\,\sin{i}$. Solid symbols are
    period measurements from \citet{2011ApJ...727...56I}, open circles
    are from \citet{Kiraga:2007p407}, open squares are taken from
    other literature (see text). Three stars with $v\,\sin{i} > v_{\rm
      eq}$ in which $v\,\sin{i}$ is not an upper limit are shown as
    red stars.}
\end{figure}

The stars with rotational period measurements are shown in
Table\,\ref{table:periods}. For each star, we calculated the
equatorial surface velocity, $v_{\rm eq}$, derived from the period and
compared it to the measured projected surface velocity
$v\,\sin{i}$. The two velocities, $v_{\rm eq}$ and $v\,\sin{i}$, are
compared in the left panel of Fig.\,\ref{fig:periods}. If period and
surface velocity are consistent, the stars should populate the region
close to the line of identity (drawn as solid line in
Fig.\,\ref{fig:periods}). Stars observed under low inclination angles
$i$ are expected to fall below that line. Comparing the values $v_{\rm
  eq}$ and $v\,\sin{i}$, we find that several stars are far away from
the line of unit slope. From the typical scatter in the mass-luminosity
relation, uncertainties in parallax and photometric measurements, and
the scatter in the mass-radius identity, we estimated that the final
uncertainty in $v_{\rm eq}$ is typically much lower than 50\,\%, which
translates into uncertainties in the inclination much lower than a
factor of 2. Therefore, very low inclination angles (below $\sim
50^{\circ}$) are unlikely to be caused by uncertainties in measuring
$v\,\sin{i}$ or the translation into $v_{\rm eq}$.  

Stars with $v\,\sin{i} < v_{\rm eq}$ may be observed under small
inclination angles, for these stars we plot inclination $i$ as a
function of rotation period in the lower right panel of
Fig.\,\ref{fig:periods}. M-type stars with rotation periods on the
order of $P = 10$\,d and longer have surface rotation velocities below
the typical detection limits of $v\,\sin{i}$ measurements. These stars
are marked with downward arrows in Fig.\,\ref{fig:periods}. Although
we cannot determine information about inclination for those stars, the
non-detection of rotational broadening means that spectroscopic
measurements are consistent with the reported photometric periods. For
some stars, however, we find measurements of rotation velocities with
$v\,\sin{i} > v_{\rm eq}$. For all stars with $v\,\sin{i} > v_{\rm
  eq}$ (including upper limits in $v\,\sin{i}$), we calculate the
ratio between projected rotation velocities and photometrically
derived surface velocity, $v\,\sin{i} / v_{\rm eq}$, and plot this
ratio in the upper right panel of Fig.\,\ref{fig:periods}. As
expected, for very long periods, limited spectral resolving power
leads to very large ratios.

The first conclusion from the comparison between photometric periods and
projected rotation velocties is that both measurements are consistent for
several stars with measured $v\,\sin{i}$ above the detection limit, and the
majority of upper limits in $v\,\sin{i}$ are consistent with surface rotation
velocities being below the spectroscopic detection limit. There are two groups
of stars in which spectroscopic and photometric rotation rates are not
consistent: (1) Among the stars with rotation periods shorter than 10\,d,
seven stars have inclination angles below $i = 60^{\circ}$ while eight stars
have larger inclination angles but ratios of $v\,\sin{i} / v_{\rm eq}$ not
much larger than 1. The fraction of stars with $i < 60^{\circ}$ is 47\,\%,
which is consistent with the assumption of random orientation of the rotation
axis (leading to an expected fraction of 50\,\% with $i < 60^{\circ}$). On the
other hand, several stars have extremely small inclination angles, for example
two stars have $i < 5^{\circ}$. The fraction of stars with such low an
inclination angle in a sample of randomly oriented spin axes is only 0.4\,\%,
yet 14\,\% of the stars in this subsample are found. Thus, the fraction of
stars with very low inclination angles appears to be unrealistically low.  (2)
Three stars exist in our sample in which $v\,\sin{i}$ exceeds $v_{\rm eq}$ by
a factor of 2 or higher (marked as red stars in
Fig.\,\ref{fig:periods}). Here, spectroscopic and photometric measurements are
clearly inconsistent.

Inconsistencies between spectroscopic and photometric measurements can have
several reasons. First, a high frequency of stars observed under very small
inclination angles could be due to an observational bias. Photometric periods
are most likely to be detected in stars that show large brightness
variations. If a star is observed pole-on, brightness variations caused by
corotating spots are smaller than if the star is observed under high
inclination. This results in a potential bias towards the detection of
photometric periods in stars with large values of $i$. Thus, this bias results
in a \emph{lower} fraction of stars with very small $i$ than expected from
random distribution of rotation axes. Taking this bias into account, the
existence of several stars with very small inclination angles is even more
unlikely. Another potential source of error are incorrect period measurements.
Period measurements from \citet{2011ApJ...727...56I} are based on several
hundred data points and show clear periodicity as demonstrated in that
paper. The quality of other period reports is generally lower simply because
of the exquisite data quality used in \citet{2011ApJ...727...56I}. For
example, photometric data used for the \citet{Kiraga:2007p407} periods are of
much lower quality. Nevertheless, many periods reported in
\citet{Kiraga:2007p407} look rather convincing as demonstrated by the authors.
In our sample, we identified five period measurements that are inconsistent
with $v\,\sin{i}$ measurements possibly because of period misidentifications;
Gl~431 and GJ~2036A have periods much lower than expected from $v\sin{i}$
($v\,\sin{i} / v_{\rm eq} \ge 2$), and Steph-546A, Wo~9520, and Gl 669A have
extremely low inclination angles ($i < 15^{\circ}$). The periods for all these
stars are from \citet{Kiraga:2007p407}, and visual inspection of their
phase-folded lightcurves indeed shows that misidentification of these periods
is likely. It is interesting to note that for GJ~2036A surface equatorial
velocity and spectroscopic projected rotation velocity differ by exactly a
factor of two. A photometric period of 1.6\,d instead of 0.8\,d would agree
with the spectroscopic measurement. A potential reason for this difference may
be that the 0.8\,d period is an alias of a 1.6\,d period.

A third potential reason for inconsistencies between spectroscopic and
photometric measurements is an incorrect estimate of $v\,\sin{i}$. This is a
likely explanation for one of our cases: GJ~1186 has $v\,\sin{i} =
3.9$\,km\,s$^{-1}$ exceeding $v_{\rm eq}$ by a factor of 40. Here, the
phase-folded lightcurve presented by \citet{2011ApJ...727...56I} looks very
reasonable. The value of $v\,\sin{i}$ is from the catalogue of
\citet{Jenkins:2009p2650} and similar to the values shown in
Table\,\ref{table:rapid_inactive} for which we argued that these measurements
are upper limits rather than detections. We argue that this measurement is
probably an upper limit, too. This point is strengthened by the fact that the
period measurements from the catalogue of \citet[][solid points in
Fig.\,\ref{fig:periods}]{2011ApJ...727...56I} seem to be rather robust -- all
periods except for GJ~1186 are consistent with $v\,\sin{i}$ data.

\section{Summary}
\label{sec:summary}

We presented a comprehensive catalogue of projected rotation velocities in
presumably single stars of spectral type M0.0--M4.5 from high-resolution
spectroscopy using 206 new spectra taken for this project and literature
values from field stars from several earlier collections. Where available, we
add information on chromospheric emission from H$\alpha$. Our catalogue
presents statistically meaningful information on rotation and activity in
stars close to the transition between partial and complete convection.

In addition, we identified 12 spectroscopic binaries plus 8 binaries that were
already known from earlier work and re-observed for our project. 13 other
stars were found to show peculiar line profiles perhaps due to binarity. These
33 stars are not contained in the catalogue and presented individually.

We investigated rotation and activity in our sample stars with an
emphasis on the transition from partial to convection convection
occuring around spectral type M3. We confirm that the fraction of
active stars is very low in early-M stars ($<$M3) and rises steeply
around spectral type M3. For the seven active early-M field stars in
our sample, we found evidence that all are younger than a few hundred
Myr. Furthermore, find that the behavior of the fraction of rapidly
rotating stars with respect to spectral class is virtually identical
to the fraction of active stars, which provides strong support to the
assumption that all active stars are rapid rotators. Detailed analysis
of the rotation-activity relation supports this picture. We argue that
in a few individual cases reports of rotational broadening in the
absence of H$\alpha$ emission are spurious and that these detections
are in fact upper limits of $v\,\sin{i}$. We conclude that all rapid
rotators ($v\,\sin{i} > 3$\,km\,s$^{-1}$) are active ($\log{L_{\rm
    H_{\alpha}}/L_{\rm bol}} > -4.5$). There is no significant
difference between rotation-activity relations on both sides of the
convection boundary. An important result is that the distribution of
activity in early- to mid-M dwarfs can entirely be explained by
rotational braking. This implies that at the boundary to complete
convection, we do not observe any evidence for a transition from a
rotationally dominated dynamo to a turbulent dynamo independent of
rotation. This does not imply that the predominant dynamo mechanism
does not change, but it shows that the dynamo in fully convective
stars at spectral type M3.0--M4.5 is still driven by rotation.

Scatter in the rotation-activity diagram appears to be different between
early-M and later stars (M3.0--M4.5). Early-M stars show less scatter in
activity while large scatter is observed in the later ones. The difference,
however, is statistically not well defined because the early-M sample consists
of five detections in $v\,\sin{i}$ only, and the distribution of the slowest
rotators among the most active stars among the later-M sample is consistent
with random distibution of rotation axes. Some stars also may have been
observed during short-time flares.

We compared projected rotation velocities to photometric periods taken from
several catalogues. Inclination angles of a few rapid rotators are reported:
most of the stars with rotation periods longer than $P = 10$\,d have rotation
velocities $v\,\sin{i}$ below our detection limit and are consistent with very
slow rotation. We identified a few cases where $v\,\sin{i}$ and $P$ are
inconsistent, in four or five cases the rotation periods are probably
mis-identifications, in one case the $v\,\sin{i}$ measurement probably is an
upper limit rather than a detection.

Our catalogue presents a comprehensive database for understanding the
evolution of low-mass stars and the connection between rotation and magnetic
activity. The latter is considered as an important factor for the development
of life on habitable planets, for which early-M dwarfs have become a prominent
target sample. The potential to detect extrasolar planets depends on the width
of the stellar line profiles, and our catalogue provides important input
selecting target samples for future radial velocity surveys for planets around
low-mass stars.

\acknowledgements Based on observations collected at the Centro
Astron\'omico Hispano Alem\`an (CAHA) at Calar Alto, operated jointly
by the Max-Planck Institut f\"ur Astronomie and the Instituto de
Astrof\'isica de Andaluc\'ia (CSIC), and on observations obtained from
the European Southern Observatory on MPI time under PID 076.A-9005. AR
acknowledges financial support from the Deutsche
Forschungsgemeinschaft (DFG) under an Emmy Noether fellowship (RE
1664/4-1) and a Heisenberg Professorship (RE 1664/9-1). NJ
acknowledges the support from the DFG Research Training Group GrK-1351
``Extrasolar Planets and their Host Stars''. This work was supported
by Sonderforschungsbereich SFB\,881 ``The Milky Way System''
(subproject B6) of the DFG. We thank an anonymous referee for a timely
and very useful report.

Facilities: \facility{La Silla(FEROS)}, \facility{Calar Alto(FOCES)}

\bibliographystyle{apj}
\bibliography{rotactpap}

\begin{deluxetable}{lcrcrrrcrr}
  \tablewidth{0pt}
  \tablecaption{\label{table:periods}Stars with photometric periods}
  \tablehead{Name & Spectral type & $P$ & ref & $v_{\rm eq}$ &
    \multicolumn{2}{c}{$v\sin i$} & ref & $i$ & exceed\\
  &&[d]&&[km\,s$^{-1}$]&\multicolumn{2}{c}{[km\,s$^{-1}$]}&&[$^\circ$]}
  \startdata
  \input{Periods_table.tex}
  \enddata
  \tablerefs{Period references: (ks) \citet{Kiraga:2007p407}; (fh)
    \citet{2000AJ....120.3265F}; (no) \citet{Noyes:1984p48}; (al)
    \citet{1998ARep...42..649A}; (be) \citet{1998AJ....116..429B};
    (ir) \citet{2011ApJ...727...56I}; (en)
    \citet{2009AIPC.1135..221E}; (co) \citet{1995A&A...300..819C} --
    $v\,\sin{i}$ references: (1) This work (2)
    \citet{Browning:2010p2615}; (3) \citet{Marcy:1992p2101}; (4)
    \citet{Reiners:2007p18}; (5) \citet{Delfosse:1998p79}; (6)
    \citet{Jenkins:2009p2650}; (7) \citet{Reiners:2007p19}; (8)
    \citet{2008A&A...489L..45R}}
\end{deluxetable}

\clearpage

\LongTables

\begin{deluxetable*}{lrrcrlrrc}
  \tablewidth{0pt}
  \tablecaption{\label{table:objects}Catalogue of ration and activity in 334 M0--M4.5 stars}
  \tablehead{Name & $\alpha (J2000)$ & $\delta (J2000)$ & Spectral type &
    \multicolumn{2}{c}{$\log({L_{\rm H_{\alpha}}/L_{\rm bol})}$} &
    \multicolumn{2}{c}{$v\sin i$} & ref\\
  &&&&&& \multicolumn{2}{c}{[km\,s$^{-1}$]}}
  \startdata
  \input{Full_table.tex}
  \enddata
  \tablerefs{\footnotesize (1) This work (2) \cite{Browning:2010p2615}; (3)
  \cite{Marcy:1992p2101}; (4) \cite{Reiners:2007p18}; (5)
  \cite{Delfosse:1998p79}; (6) \cite{Jenkins:2009p2650}; (7)
  \cite{Reiners:2007p19}}
\end{deluxetable*}

\clearpage

\begin{deluxetable}{lcccccccccc}
  \tablewidth{0pt} \tablecaption{\label{tab:vsinicomp}Stars with
    rotation measurements from different sources. Uncertainties are
    discussed in Sects.\,\ref{ssec:vsini_det} and \ref{ssec:rot_cs}.}
  \tablehead{Name & Spectral type & \multicolumn{7}{c}{$v\sin i$ [km\,s$^{-1}$]}\\
    && this work & (1) & (2) & (3) & (4) & (5) & (6) & adopted}
  \startdata
  \input{vsinicomp_table.tex}
  \enddata
  \tablerefs{\footnotesize (1) \citet{Delfosse:1998p79}; (2)
    \citet{Jenkins:2009p2650}; (3) \citet{Marcy:1992p2101}; (4)
    \citet{Reiners:2007p18}; (5) \citet{Reiners:2007p19}; (6)
    \citet{Browning:2010p2615}}
\end{deluxetable}

\end{document}

%% file: EarlyActive_table.tex
          Gl 182 & M0.0 &     & -4.11 &     &  10.4 \\
          Gl 494 & M0.5 &     & -3.80 &     &   9.7 \\
      Steph 546A & M1.5 &     & -3.89 &     &   5.3 \\
         Wo 9520 & M1.5 &     & -3.92 &     &   6.5 \\
        GJ 2036A & M2.0 &     & -3.45 &     &  44.3 \\
          Gl 358 & M2.0 &     & -4.44 & $<$ &   3.0 \\
         Gl 569A & M2.0 &     & -4.30 & $<$ &   2.5 \\

%% file: FastInactive_table.tex
         GJ 3104 & M3.0 & $<$ & -4.90* &     &   4.0 \\
        LHS 2651 & M3.5 & $<$ &  -4.93 &     &   3.9 \\
        LHS 1785 & M4.5 & $<$ &  -4.97 &     &   4.5 \\
        LHS 1857 & M4.5 & $<$ &  -4.97 &     &   4.0 \\
         GJ 3542 & M4.5 & $<$ &  -4.97 &     &   3.9 \\
         GJ 1134 & M4.5 & $<$ & -4.97* &     &   4.1 \\
       G 121-028 & M4.5 & $<$ & -4.97* &     &   3.8 \\
         GJ 1186 & M4.5 & $<$ &  -4.97 &     &   3.9 \\
          Gl 585 & M4.5 & $<$ & -4.97* &     &   3.1 \\

%% file: Periods_table.tex
          Gl 182 & M0.0 &   4.4 & (ks) &   7.9 &     &  10.4 & (1) &    &  1.3 \\
          Gl 410 & M0.0 &  14.8 & (fh) &   2.0 & $<$ &   2.5 & (2) &    &      \\
          Gl 424 & M0.0 & 149.7 & (en) &   0.2 & $<$ &   2.5 & (2) &    &      \\
          Gl 494 & M0.5 &   2.9 & (ks) &   9.8 &     &   9.7 & (2) & 81 &      \\
      Steph 546A & M1.5 &   0.3 & (ks) &  74.2 &     &   5.3 & (1) &  4 &      \\
          Gl 205 & M1.5 &  33.6 & (ks) &   0.9 &     &   1.5 & (4) &    &  1.7 \\
          Gl 382 & M1.5 &  21.6 & (ks) &   1.2 &     &   1.8 & (4) &    &  1.5 \\
         Wo 9520 & M1.5 &   0.4 & (ks) &  69.9 &     &   6.5 & (1) &  5 &      \\
        GJ 2036A & M2.0 &   0.8 & (ks) &  22.6 &     &  44.3 & (1) &    &  2.0 \\
          Gl 358 & M2.0 &  25.3 & (ks) &   0.8 & $<$ &   3.0 & (1) &    &      \\
          Gl 411 & M2.0 &  48.0 & (no) &   0.5 & $<$ &   2.5 & (2) &    &      \\
         Gl 569A & M2.0 &  13.7 & (ks) &   1.7 & $<$ &   2.5 & (2) &    &      \\
           Gl 84 & M2.5 &  44.5 & (ks) &   0.5 & $<$ &   3.0 & (1) &    &      \\
          Gl 674 & M2.5 &  33.3 & (ks) &   0.6 & $<$ &   3.0 & (1) &    &      \\
          Gl 388 & M3.0 &   2.2 & (en) &   8.2 &     &   3.0 & (4) & 21 &      \\
          Gl 735 & M3.0 &   2.9 & (al) &   9.0 &     &   7.7 & (1) & 58 &      \\
          Gl 431 & M3.5 &  14.3 & (ks) &   1.2 &     &  20.5 & (1) &    & 17.6 \\
         Gl 669A & M3.5 &   0.9 & (ks) &  18.6 & $<$ &   4.0 & (1) & 12 &      \\
          Gl 729 & M3.5 &   2.9 & (ks) &   3.5 &     &   4.0 & (2) &    &  1.1 \\
          Gl 873 & M3.5 &   4.4 & (co) &   3.4 &     &   3.5 & (2) &    &  1.0 \\
       G 099-049 & M4.0 &   1.8 & (ir) &   7.3 &     &   7.4 & (5) &    &  1.0 \\
          Gl 699 & M4.0 & 130.0 & (be) &   0.1 & $<$ &   2.5 & (2) &    &      \\
         GJ 1243 & M4.0 &   0.6 & (ir) &  22.2 &     &  22.0 & (1) & 83 &      \\
          Gl 876 & M4.0 & 116.5 & (en) &   0.1 & $<$ &   2.5 & (2) &    &      \\
        LHS 1885 & M4.5 &  52.4 & (ir) &   0.2 & $<$ &   3.7 & (5) &    &      \\
          Gl 285 & M4.5 &   2.8 & (ir) &   5.8 &     &   4.5 & (4) & 50 &      \\
         GJ 1151 & M4.5 & 132.0 & (ir) &   0.1 & $<$ &   4.1 & (5) &    &      \\
        Gl 493.1 & M4.5 &   0.6 & (ir) &  16.0 &     &  16.8 & (5) &    &  1.0 \\
         GJ 1186 & M4.5 &  88.3 & (ir) &   0.1 &     &   3.9 & (6) &    & 40.1 \\
        Gl 791.2 & M4.5 &   0.3 & (ir) &  33.6 &     &  32.0 & (5) & 72 &      \\
\hline
         GJ 1057 & M5.0 & 102.0 & (ir) &   0.1 & $<$ &   2.2 & (5) &    &      \\
         GJ 1156 & M5.0 &   0.5 & (ir) &  16.5 &     &   9.2 & (5) & 33 &      \\
          Gl 551 & M5.5 &  82.5 & (ks) &   0.1 & $<$ &   3.0 & (8) &    &      \\

%% file: Full_table.tex
         LTT 692 & 01 14 33.9 & $-$53 56 39 & M0.0 & $<$ &  -4.82 & $<$ &   3.0 & (1) \\
       LTT 11085 & 03 18 38.1 &    32 39 57 & M0.0 & $<$ &  -4.82 & $<$ &   4.0 & (1) \\
          Gl 182 & 04 59 34.7 &    01 47 00 & M0.0 &     &  -4.11 &     &  10.4 & (1) \\
          Gl 353 & 09 31 56.4 &    36 19 16 & M0.0 &     &     -- & $<$ &   2.5 & (2) \\
          Gl 373 & 09 56 08.9 &    62 47 21 & M0.0 &     &     -- & $<$ &   2.5 & (2) \\
          Gl 410 & 11 02 38.2 &    21 58 01 & M0.0 &     &     -- & $<$ &   2.5 & (2) \\
          Gl 424 & 11 19 57.7 &    65 50 33 & M0.0 &     &     -- & $<$ &   2.5 & (2) \\
          Gl 438 & 11 43 18.1 & $-$51 50 14 & M0.0 & $<$ &  -4.82 & $<$ &   3.0 & (1) \\
        Gl 459.3 & 12 19 24.4 &    28 22 55 & M0.0 &     &     -- & $<$ &   3.0 & (3) \\
         Gl 461A & 12 20 25.4 &    00 34 59 & M0.0 &     &     -- & $<$ &   3.0 & (3) \\
         Gl 548A & 14 25 42.9 &    23 37 10 & M0.0 & $<$ &  -4.82 & $<$ &   4.0 & (1) \\
           V 759 & 16 09 02.9 &    52 56 36 & M0.0 & $<$ &  -4.82 & $<$ &   4.0 & (1) \\
           V 791 & 17 18 21.6 & $-$01 46 51 & M0.0 & $<$ &  -4.82 & $<$ &   4.0 & (1) \\
         Gl 676A & 17 30 11.6 & $-$51 38 11 & M0.0 & $<$ &  -4.82 & $<$ &   3.0 & (1) \\
       Gl 678.1A & 17 30 22.6 &    05 32 55 & M0.0 & $<$ &  -4.82 & $<$ &   2.5 & (2) \\
        Gl 694.2 & 17 45 33.4 &    46 51 18 & M0.0 & $<$ &  -4.82 & $<$ &   4.0 & (1) \\
          Gl 696 & 17 50 33.9 & $-$06 02 59 & M0.0 & $<$ &  -4.82 & $<$ &   4.0 & (1) \\
       G 183-041 & 18 25 04.7 &    24 38 08 & M0.0 & $<$ &  -4.82 & $<$ &   4.0 & (1) \\
         Gl 720A & 18 35 18.0 &    45 44 35 & M0.0 & $<$ &  -4.82 & $<$ &   3.0 & (3) \\
          Gl 731 & 18 51 51.2 &    16 35 03 & M0.0 & $<$ &  -4.82 & $<$ &   4.0 & (1) \\
       G 141-052 & 18 59 38.3 &    07 59 14 & M0.0 & $<$ &  -4.82 & $<$ &   4.0 & (1) \\
          V 811A & 19 35 06.3 &    08 27 39 & M0.0 & $<$ &  -4.82 & $<$ &   4.0 & (1) \\
       G 260-030 & 19 39 33.1 &    71 52 22 & M0.0 & $<$ &  -4.82 & $<$ &   4.0 & (1) \\
       G 210-045 & 20 58 41.7 &    34 16 27 & M0.0 & $<$ &  -4.82 & $<$ &   4.0 & (1) \\
          Gl 835 & 21 38 00.0 &    27 43 25 & M0.0 & $<$ &  -4.82 & $<$ &   4.0 & (1) \\
       Gl 838.3B & 21 51 53.3 &    42 20 39 & M0.0 & $<$ &  -4.82 & $<$ &   4.0 & (1) \\
          Gl 839 & 21 53 58.7 &    41 46 46 & M0.0 & $<$ &  -4.82 & $<$ &   4.0 & (1) \\
          Gl 846 & 22 02 10.4 &    01 24 02 & M0.0 & $<$ &  -4.82 & $<$ &   2.5 & (2) \\
         Wo 9784 & 22 28 45.8 &    18 55 54 & M0.0 & $<$ &  -4.82 & $<$ &   4.0 & (1) \\
       LHS 1051A & 00 15 49.3 & $-$67 59 49 & M0.5 & $<$ &  -4.87 & $<$ &   3.0 & (1) \\
         GJ 2003 & 00 20 08.2 & $-$17 03 38 & M0.5 & $<$ &  -4.87 & $<$ &   3.0 & (1) \\
           Gl 21 & 00 26 52.9 &    70 08 33 & M0.5 & $<$ &  -4.87 & $<$ &   4.0 & (1) \\
         Gl 27.1 & 00 39 57.8 & $-$44 15 08 & M0.5 & $<$ &  -4.87 & $<$ &   3.0 & (2) \\
        Gl 84.1A & 02 05 23.3 & $-$28 04 16 & M0.5 & $<$ &  -4.87 & $<$ &   3.0 & (1) \\
        Gl 155.1 & 03 47 58.2 &    02 47 18 & M0.5 & $<$ &  -4.87 & $<$ &   3.0 & (1) \\
         GJ 1074 & 04 58 45.5 &    50 56 39 & M0.5 & $<$ &  -4.87 & $<$ &   4.0 & (1) \\
          Gl 184 & 05 03 22.7 &    53 07 55 & M0.5 & $<$ &  -4.87 & $<$ &   4.0 & (1) \\
          Gl 212 & 05 41 30.7 &    53 29 27 & M0.5 & $<$ &  -4.87 & $<$ &   2.5 & (2) \\
          Gl 229 & 06 10 34.6 & $-$21 51 46 & M0.5 & $<$ &  -4.87 &     &   1.0 & (4) \\
        Gl 277.1 & 07 34 28.1 &    62 56 30 & M0.5 &     &     -- & $<$ &   2.5 & (2) \\
        Gl 336.1 & 09 11 31.0 &    46 37 01 & M0.5 & $<$ &  -4.87 & $<$ &   4.0 & (1) \\
          Gl 369 & 09 51 08.9 & $-$12 19 34 & M0.5 & $<$ &  -4.87 & $<$ &   3.0 & (1) \\
         Gl 412A & 11 05 24.6 &    43 31 41 & M0.5 &     &     -- & $<$ &   2.5 & (2) \\
          Gl 494 & 13 00 46.8 &    12 22 32 & M0.5 &     &  -3.80 &     &   9.7 & (2) \\
         Gl 507A & 13 19 33.3 &    35 06 41 & M0.5 & $<$ &  -4.87 & $<$ &   4.0 & (1) \\
          Gl 514 & 13 29 59.0 &    10 22 46 & M0.5 & $<$ &  -4.91 &     &   1.5 & (4) \\
           V 150 & 14 02 19.5 &    13 41 24 & M0.5 & $<$ &  -4.87 & $<$ &   4.0 & (1) \\
         Gl 537A & 14 02 32.7 &    46 20 24 & M0.5 & $<$ &  -4.87 & $<$ &   4.0 & (1) \\
         Gl 537B & 14 02 32.7 &    46 20 24 & M0.5 & $<$ &  -4.87 & $<$ &   4.0 & (1) \\
         Gl 548B & 14 25 46.1 &    23 37 22 & M0.5 & $<$ &  -4.87 & $<$ &   4.0 & (1) \\
      Steph 1453 & 17 15 50.2 &    18 59 58 & M0.5 & $<$ &  -4.87 & $<$ &   4.0 & (1) \\
          Gl 685 & 17 35 34.0 &    61 40 57 & M0.5 & $<$ &  -4.87 & $<$ &   4.0 & (1) \\
       G 182-037 & 18 04 17.5 &    35 57 27 & M0.5 & $<$ &  -4.87 & $<$ &   4.0 & (1) \\
          Gl 709 & 18 16 31.0 &    45 33 24 & M0.5 & $<$ &  -4.87 & $<$ &   4.0 & (1) \\
          Gl 740 & 18 58 00.1 &    05 54 36 & M0.5 & $<$ &  -4.87 & $<$ &   4.0 & (1) \\
          Gl 809 & 20 53 19.7 &    62 09 20 & M0.5 &     &     -- & $<$ &   2.5 & (2) \\
        Gl 842.2 & 21 58 24.2 &    75 35 19 & M0.5 & $<$ &  -4.87 & $<$ &   4.0 & (1) \\
          Gl 887 & 23 05 52.0 & $-$35 51 12 & M0.5 &     &     -- & $<$ &   2.5 & (2) \\
            Gl 2 & 00 05 10.2 &    45 47 12 & M1.0 & $<$ &  -4.88 & $<$ &   2.5 & (2) \\
          Gl 15A & 00 18 20.8 &    44 01 19 & M1.0 &     &     -- & $<$ &   2.5 & (2) \\
       G 036-038 & 02 52 24.9 &    26 58 31 & M1.0 & $<$ &  -4.88 & $<$ &   4.0 & (1) \\
         Gl 119A & 02 56 33.7 &    55 26 15 & M1.0 & $<$ &  -4.88 & $<$ &   4.0 & (1) \\
       G 246-026 & 03 10 26.4 &    58 26 08 & M1.0 & $<$ &  -4.88 & $<$ &   4.0 & (1) \\
         Gl 140A & 03 24 06.4 &    23 47 06 & M1.0 & $<$ &  -4.88 & $<$ &   4.0 & (1) \\
       Gl 150.1B & 03 43 45.1 &    16 40 03 & M1.0 & $<$ &  -4.88 & $<$ &   3.0 & (1) \\
        Wo 9163A & 04 40 29.2 & $-$09 11 44 & M1.0 & $<$ &  -4.88 & $<$ &   3.0 & (1) \\
       G 081-036 & 04 50 15.1 &    45 58 51 & M1.0 & $<$ &  -4.88 & $<$ &   4.0 & (1) \\
          Gl 367 & 09 44 30.5 & $-$45 46 25 & M1.0 & $<$ &  -4.88 & $<$ &   3.0 & (1) \\
          Gl 390 & 10 25 11.0 & $-$10 13 44 & M1.0 & $<$ &  -4.88 & $<$ &   2.5 & (2) \\
          Gl 450 & 11 51 07.5 &    35 16 16 & M1.0 &     &     -- & $<$ &   2.5 & (2) \\
          Gl 477 & 12 35 58.6 & $-$45 56 05 & M1.0 & $<$ &  -4.88 & $<$ &   3.0 & (1) \\
          Gl 521 & 13 39 24.1 &    46 11 07 & M1.0 &     &     -- & $<$ &   3.0 & (3) \\
          Gl 536 & 14 01 03.7 & $-$02 39 23 & M1.0 & $<$ &  -4.88 & $<$ &   2.5 & (2) \\
         Gl 570B & 14 57 27.3 & $-$21 25 02 & M1.0 &     &     -- & $<$ &   2.5 & (2) \\
          Gl 606 & 15 59 53.1 & $-$08 15 11 & M1.0 & $<$ &  -4.88 & $<$ &   3.0 & (1) \\
        LP 806-8 & 16 48 45.8 & $-$15 44 17 & M1.0 & $<$ &  -4.88 & $<$ &   4.0 & (1) \\
          Gl 649 & 16 58 08.8 &    25 44 41 & M1.0 & $<$ &  -4.88 & $<$ &   2.5 & (2) \\
       G 139-023 & 17 16 00.6 &    11 03 29 & M1.0 & $<$ &  -4.88 & $<$ &   4.0 & (1) \\
          Gl 686 & 17 37 52.7 &    18 35 21 & M1.0 &     &     -- & $<$ &   2.5 & (2) \\
          Gl 701 & 18 05 07.0 & $-$03 01 47 & M1.0 &     &     -- & $<$ &   2.5 & (2) \\
          Gl 724 & 18 40 57.3 & $-$13 22 42 & M1.0 & $<$ &  -4.88 & $<$ &   4.0 & (1) \\
       G 125-030 & 19 45 49.4 &    32 23 10 & M1.0 & $<$ &  -4.88 & $<$ &   4.0 & (1) \\
         Gl 767A & 19 46 23.7 &    32 01 02 & M1.0 & $<$ &  -4.88 & $<$ &   4.0 & (1) \\
       G 262-029 & 20 51 59.7 &    69 10 07 & M1.0 & $<$ &  -4.88 & $<$ &   4.0 & (1) \\
          Gl 821 & 21 09 17.0 & $-$13 17 54 & M1.0 & $<$ &  -4.88 & $<$ &   2.5 & (2) \\
          Gl 863 & 22 33 01.9 &    09 22 40 & M1.0 &     &     -- & $<$ &   3.0 & (3) \\
          Gl 895 & 23 24 30.4 &    57 51 17 & M1.0 &     &     -- & $<$ &   2.5 & (2) \\
          Gl 908 & 23 49 11.9 &    02 24 11 & M1.0 &     &     -- & $<$ &   2.5 & (2) \\
            Gl 1 & 00 05 24.4 & $-$37 21 25 & M1.5 & $<$ &  -4.88 & $<$ &   2.5 & (2) \\
      G 242-048A & 00 13 39.9 &    80 39 26 & M1.5 & $<$ &  -4.88 & $<$ &   4.0 & (1) \\
           Gl 16 & 00 18 16.5 &    10 12 09 & M1.5 & $<$ &  -4.88 & $<$ &   3.0 & (1) \\
         GJ 1009 & 00 21 55.8 & $-$31 24 18 & M1.5 & $<$ &  -4.88 & $<$ &   2.5 & (2) \\
           Gl 49 & 01 02 38.0 &    62 20 40 & M1.5 &     &     -- & $<$ &   3.4 & (5) \\
           Gl 87 & 02 12 21.8 &    03 34 45 & M1.5 & $<$ &  -4.88 & $<$ &   2.5 & (2) \\
           Gl 91 & 02 13 52.3 & $-$32 02 19 & M1.5 & $<$ &  -4.88 & $<$ &   3.0 & (1) \\
         GJ 1051 & 02 43 52.6 & $-$08 49 36 & M1.5 & $<$ &  -4.88 & $<$ &   3.0 & (1) \\
       Gl 114.1A & 02 50 03.6 & $-$53 08 36 & M1.5 & $<$ &  -4.88 & $<$ &   3.0 & (1) \\
       Gl 156.1A & 03 56 47.1 &    53 33 39 & M1.5 & $<$ &  -4.88 & $<$ &   4.0 & (1) \\
          Gl 173 & 04 37 41.9 & $-$11 02 18 & M1.5 & $<$ &  -4.88 & $<$ &   2.5 & (2) \\
      Steph 546A & 05 06 49.6 & $-$21 35 06 & M1.5 &     &  -3.89 &     &   5.3 & (1) \\
          Gl 205 & 05 31 26.9 & $-$03 40 22 & M1.5 & $<$ &  -4.88 &     &   1.5 & (4) \\
          Gl 218 & 05 47 39.3 & $-$36 19 41 & M1.5 & $<$ &  -4.88 & $<$ &   3.0 & (1) \\
        LTT 3412 & 09 16 20.9 & $-$18 37 35 & M1.5 & $<$ &  -4.88 & $<$ &   3.0 & (1) \\
          Gl 361 & 09 41 10.6 &    13 12 34 & M1.5 &     &     -- & $<$ &   2.5 & (2) \\
          Gl 382 & 10 12 17.6 & $-$03 44 42 & M1.5 & $<$ &  -4.89 &     &   1.8 & (4) \\
         Gl 414B & 11 11 02.2 &    30 26 42 & M1.5 &     &     -- & $<$ &   2.5 & (2) \\
       Steph 928 & 11 12 38.9 &    18 56 04 & M1.5 & $<$ &  -4.88 & $<$ &   3.0 & (1) \\
          Gl 433 & 11 35 27.1 & $-$32 32 08 & M1.5 &     &     -- & $<$ &   2.5 & (2) \\
         GJ 2097 & 13 07 04.3 &    20 48 38 & M1.5 &     &     -- & $<$ &   3.7 & (5) \\
        Gl 507.1 & 13 19 40.2 &    33 20 47 & M1.5 &     &     -- & $<$ &   3.0 & (3) \\
          Gl 526 & 13 45 42.8 &    14 53 39 & M1.5 & $<$ &  -5.02 &     &   2.0 & (4) \\
         Wo 9492 & 14 42 21.8 &    66 03 19 & M1.5 & $<$ &  -4.88 & $<$ &   2.5 & (2) \\
         Wo 9520 & 15 21 52.8 &    20 58 38 & M1.5 &     &  -3.92 &     &   6.5 & (1) \\
          Gl 625 & 16 25 24.1 &    54 18 15 & M1.5 &     &     -- & $<$ &   2.5 & (2) \\
         Gl 667C & 17 18 57.7 & $-$34 59 46 & M1.5 & $<$ &  -4.88 & $<$ &   2.5 & (2) \\
          Gl 680 & 17 35 13.6 & $-$48 40 55 & M1.5 & $<$ &  -4.88 & $<$ &   3.0 & (1) \\
         Gl 745A & 19 07 05.8 &    20 53 18 & M1.5 &     &     -- & $<$ &   2.5 & (2) \\
         Gl 800A & 20 42 56.5 & $-$18 54 54 & M1.5 & $<$ &  -4.88 & $<$ &   4.0 & (1) \\
          Gl 806 & 20 45 03.6 &    44 29 44 & M1.5 & $<$ &  -4.88 & $<$ &   2.5 & (2) \\
          Gl 880 & 22 56 35.3 &    16 33 13 & M1.5 &     &     -- & $<$ &   2.5 & (2) \\
       Gl 22A.01 & 00 32 27.3 &    67 14 09 & M2.0 & $<$ &  -4.88 & $<$ &   4.0 & (1) \\
           Gl 47 & 01 01 19.6 &    61 22 02 & M2.0 & $<$ &  -4.88 & $<$ &   4.0 & (1) \\
        GJ 1026A & 01 03 13.7 &    20 05 51 & M2.0 & $<$ &  -4.88 & $<$ &   3.0 & (1) \\
         GJ 1030 & 01 06 41.5 &    15 16 22 & M2.0 & $<$ &  -4.88 & $<$ &   4.0 & (1) \\
           Gl 70 & 01 43 20.3 &    04 19 23 & M2.0 & $<$ &  -4.88 & $<$ &   2.5 & (2) \\
           Gl 78 & 01 51 48.3 & $-$10 48 08 & M2.0 & $<$ &  -4.88 & $<$ &   3.0 & (1) \\
          Gl 133 & 03 21 20.5 &    79 57 59 & M2.0 & $<$ &  -4.88 & $<$ &   4.0 & (1) \\
        GJ 2036A & 04 53 30.8 & $-$55 51 34 & M2.0 &     &  -3.45 &     &  44.3 & (1) \\
          Gl 180 & 04 53 49.6 & $-$17 46 16 & M2.0 & $<$ &  -4.88 & $<$ &   2.5 & (2) \\
          Gl 192 & 05 12 42.0 &    19 39 53 & M2.0 &     &     -- & $<$ &   2.5 & (2) \\
         GJ 1077 & 05 16 59.4 & $-$78 16 54 & M2.0 & $<$ &  -4.88 & $<$ &   3.0 & (1) \\
          Gl 226 & 06 10 19.3 &    82 06 33 & M2.0 & $<$ &  -4.88 & $<$ &   2.5 & (2) \\
         Gl 250B & 06 52 18.3 & $-$05 11 22 & M2.0 &     &     -- & $<$ &   2.5 & (2) \\
         GJ 2066 & 08 16 08.1 &    01 18 07 & M2.0 &     &     -- & $<$ &   2.5 & (2) \\
          Gl 358 & 09 39 47.3 & $-$41 04 11 & M2.0 &     &  -4.44 & $<$ &   3.0 & (1) \\
          Gl 372 & 09 53 11.8 & $-$03 41 20 & M2.0 & $<$ &  -4.88 & $<$ &   4.0 & (1) \\
          Gl 393 & 10 28 55.8 &    00 50 32 & M2.0 & $<$ &  -4.83 &     &   1.5 & (4) \\
          Gl 411 & 11 03 22.3 &    35 57 20 & M2.0 &     &     -- & $<$ &   2.5 & (2) \\
        Gl 413.1 & 11 09 32.1 & $-$24 35 49 & M2.0 & $<$ &  -4.88 & $<$ &   2.5 & (2) \\
          Gl 465 & 12 24 51.8 & $-$18 14 16 & M2.0 & $<$ &  -4.88 & $<$ &   2.5 & (2) \\
          Gl 479 & 12 37 54.4 & $-$52 00 06 & M2.0 & $<$ &  -4.88 & $<$ &   3.0 & (1) \\
          Gl 552 & 14 29 30.2 &    15 31 46 & M2.0 &     &     -- & $<$ &   2.5 & (2) \\
         Gl 569A & 14 54 29.0 &    16 06 05 & M2.0 &     &  -4.30 & $<$ &   2.5 & (2) \\
         Gl 618A & 16 20 04.9 & $-$37 32 08 & M2.0 & $<$ &  -4.88 & $<$ &   3.0 & (1) \\
       LTT 14949 & 16 40 48.9 &    36 18 57 & M2.0 & $<$ &  -4.88 & $<$ &   4.0 & (1) \\
          Gl 654 & 17 05 14.1 & $-$05 05 28 & M2.0 &     &     -- & $<$ &   3.0 & (3) \\
         GJ 2128 & 17 16 41.1 &    08 03 30 & M2.0 & $<$ &  -4.88 & $<$ &   4.0 & (1) \\
          Gl 739 & 18 59 07.1 & $-$48 16 15 & M2.0 & $<$ &  -4.88 & $<$ &   3.0 & (1) \\
         Gl 745B & 19 07 13.5 &    20 52 37 & M2.0 &     &     -- & $<$ &   2.5 & (2) \\
          Gl 851 & 22 11 29.8 &    18 25 32 & M2.0 &     &     -- & $<$ &   2.5 & (2) \\
          Gl 891 & 23 10 15.0 & $-$25 55 52 & M2.0 &     &     -- & $<$ &   2.5 & (2) \\
           Gl 26 & 00 38 58.0 &    30 36 57 & M2.5 & $<$ &  -4.89 & $<$ &   2.5 & (2) \\
           Gl 63 & 01 38 22.0 &    57 13 55 & M2.5 & $<$ &  -4.89 & $<$ &   4.0 & (1) \\
       G 244-037 & 01 51 50.8 &    64 26 07 & M2.5 & $<$ &  -4.89 & $<$ &   4.0 & (1) \\
           Gl 84 & 02 05 04.1 & $-$17 36 51 & M2.5 & $<$ &  -4.89 & $<$ &   3.0 & (1) \\
         GJ 1046 & 02 19 07.7 & $-$36 46 52 & M2.5 & $<$ &  -4.89 & $<$ &   3.0 & (1) \\
          Gl 145 & 03 32 56.5 & $-$44 42 10 & M2.5 & $<$ &  -4.89 & $<$ &   3.0 & (1) \\
          Gl 238 & 06 33 51.1 & $-$58 31 58 & M2.5 & $<$ &  -4.89 & $<$ &   3.0 & (1) \\
         GJ 1099 & 07 34 17.6 &    00 59 12 & M2.5 & $<$ &  -4.89 & $<$ &   3.0 & (1) \\
          Gl 357 & 09 36 01.5 & $-$21 39 31 & M2.5 &     &     -- & $<$ &   2.5 & (2) \\
          Gl 381 & 10 12 04.3 & $-$02 41 00 & M2.5 & $<$ &  -4.89 & $<$ &   4.0 & (1) \\
          Gl 399 & 10 39 40.9 & $-$06 55 23 & M2.5 & $<$ &  -4.89 & $<$ &   4.0 & (1) \\
          Gl 408 & 11 00 04.5 &    22 50 00 & M2.5 &     &     -- & $<$ &   2.5 & (2) \\
          Gl 436 & 11 42 10.5 &    26 42 30 & M2.5 & $<$ &  -4.89 & $<$ &   2.5 & (2) \\
          Gl 476 & 12 35 00.9 &    09 49 45 & M2.5 & $<$ &  -4.89 & $<$ &   3.0 & (1) \\
          Gl 588 & 15 32 14.2 & $-$41 16 20 & M2.5 & $<$ &  -4.89 & $<$ &   3.0 & (1) \\
          Gl 623 & 16 24 08.4 &    48 21 12 & M2.5 &     &     -- & $<$ &   2.9 & (5) \\
          Gl 671 & 17 19 52.4 &    41 42 56 & M2.5 &     &     -- & $<$ &   2.5 & (2) \\
          Gl 674 & 17 28 39.2 & $-$46 53 33 & M2.5 & $<$ &  -4.89 & $<$ &   3.0 & (1) \\
          Gl 694 & 17 43 55.8 &    43 22 47 & M2.5 &     &     -- & $<$ &   2.5 & (2) \\
         GJ 2138 & 18 38 44.6 & $-$14 29 22 & M2.5 & $<$ &  -4.89 & $<$ &   3.0 & (1) \\
         Gl 752A & 19 16 54.7 &    05 09 55 & M2.5 &     &     -- & $<$ &   2.5 & (2) \\
          Gl 793 & 20 30 31.4 &    65 26 55 & M2.5 &     &     -- & $<$ &   2.5 & (2) \\
          Gl 22B & 00 32 27.3 &    67 14 09 & M3.0 & $<$ &  -4.90 & $<$ &   4.0 & (1) \\
           Gl 48 & 01 02 29.5 &    71 40 50 & M3.0 &     &     -- & $<$ &   2.5 & (2) \\
         GJ 3104 & 01 39 31.2 &    05 03 18 & M3.0 & $<$ & -4.90* &     &   4.0 & (6) \\
    G 244-047.01 & 02 01 35.6 &    63 46 11 & M3.0 & $<$ &  -4.90 & $<$ &   2.5 & (2) \\
       LHS 1377A & 02 16 40.1 & $-$30 59 23 & M3.0 &     &  -3.93 & $<$ &   3.0 & (1) \\
          Gl 109 & 02 44 14.9 &    25 31 25 & M3.0 &     &     -- & $<$ &   2.5 & (2) \\
         Gl 119B & 02 56 34.5 &    55 26 32 & M3.0 & $<$ &  -4.90 & $<$ &   4.0 & (1) \\
       LP 771-96 & 03 01 51.4 & $-$16 35 32 & M3.0 & $<$ &  -4.90 & $<$ &   3.0 & (1) \\
        LHS 1731 & 05 03 20.1 & $-$17 22 22 & M3.0 & $<$ &  -4.90 & $<$ &   2.5 & (2) \\
       G 085-041 & 05 07 49.2 &    17 58 59 & M3.0 & $<$ &  -4.90 & $<$ &   3.0 & (1) \\
          Gl 251 & 06 54 49.4 &    33 16 07 & M3.0 &     &     -- & $<$ &   2.5 & (2) \\
         Gl 257A & 06 57 48.6 & $-$44 17 26 & M3.0 & $<$ &  -4.90 & $<$ &   3.0 & (1) \\
         GJ 1097 & 07 28 45.1 & $-$03 17 45 & M3.0 &     &     -- & $<$ &   2.5 & (2) \\
        LHS 1935 & 07 38 40.7 & $-$21 13 25 & M3.0 & $<$ &  -4.90 & $<$ &   2.5 & (2) \\
          Gl 362 & 09 42 52.5 &    70 02 23 & M3.0 &     &  -4.41 & $<$ &   2.5 & (2) \\
          Gl 377 & 10 01 11.3 & $-$30 23 31 & M3.0 & $<$ &  -4.90 & $<$ &   3.0 & (1) \\
          Gl 386 & 10 16 46.1 & $-$11 57 36 & M3.0 & $<$ &  -4.90 & $<$ &   4.0 & (1) \\
          Gl 388 & 10 19 36.4 &    19 52 11 & M3.0 &     &  -3.80 &     &   3.0 & (4) \\
          Gl 443 & 11 46 42.5 & $-$14 00 43 & M3.0 & $<$ &  -4.90 & $<$ &   3.0 & (1) \\
        LTT 4562 & 12 11 11.9 & $-$19 57 34 & M3.0 & $<$ &  -4.90 & $<$ &   3.0 & (1) \\
          Gl 480 & 12 38 53.1 &    11 41 47 & M3.0 &     &     -- & $<$ &   3.0 & (3) \\
         Gl 512A & 13 28 21.0 & $-$02 21 32 & M3.0 & $<$ &  -4.90 & $<$ &   4.0 & (1) \\
        LHS 3030 & 15 09 35.9 &    03 09 56 & M3.0 & $<$ &  -4.90 & $<$ &   4.0 & (1) \\
          Gl 581 & 15 19 27.5 & $-$07 43 20 & M3.0 &     &     -- & $<$ &   2.5 & (2) \\
         Gl 617B & 16 16 45.8 &    67 15 20 & M3.0 & $<$ &  -4.90 & $<$ &   2.5 & (2) \\
          Gl 655 & 17 07 07.6 &    21 33 14 & M3.0 & $<$ &  -4.90 & $<$ &   2.5 & (2) \\
          Gl 687 & 17 36 26.3 &    68 20 30 & M3.0 &     &     -- & $<$ &   2.5 & (2) \\
        LHS 3343 & 17 57 50.9 &    46 35 14 & M3.0 & $<$ &  -4.90 & $<$ &   4.0 & (1) \\
         LHS 462 & 18 18 04.1 &    38 46 40 & M3.0 & $<$ &  -4.90 & $<$ &   2.5 & (2) \\
       G 206-040 & 18 41 59.1 &    31 49 49 & M3.0 &     &     -- & $<$ &   2.5 & (2) \\
         Gl 725A & 18 42 48.0 &    59 37 29 & M3.0 &     &     -- & $<$ &   2.5 & (2) \\
          Gl 735 & 18 55 27.3 &    08 24 09 & M3.0 &     &  -3.91 &     &   7.7 & (1) \\
       G 207-019 & 19 08 29.9 &    32 16 53 & M3.0 &     &     -- & $<$ &   2.5 & (2) \\
         Gl 860A & 22 28 00.3 &    57 41 48 & M3.0 &     &     -- & $<$ &   2.5 & (2) \\
          Gl 15B & 00 18 23.8 &    44 01 35 & M3.5 &     &     -- & $<$ &   3.1 & (5) \\
           Gl 46 & 00 58 27.0 & $-$27 51 22 & M3.5 & $<$ &  -4.93 & $<$ &   3.0 & (1) \\
       LP 771-95 & 03 01 51.4 & $-$16 35 32 & M3.5 &     &  -3.89 &     &   5.5 & (1) \\
        LHS 1513 & 03 11 33.6 & $-$38 47 18 & M3.5 & $<$ &  -4.93 & $<$ &   3.0 & (1) \\
         GJ 1065 & 03 50 44.5 & $-$06 05 30 & M3.5 & $<$ &  -4.93 & $<$ &   4.0 & (1) \\
          Gl 163 & 04 09 13.3 & $-$53 22 36 & M3.5 & $<$ &  -4.93 & $<$ &   3.0 & (1) \\
          Gl 179 & 04 52 05.6 &    06 28 37 & M3.5 & $<$ &  -4.93 & $<$ &   2.5 & (2) \\
        GJ 2036B & 04 53 30.8 & $-$55 51 34 & M3.5 &     &  -3.52 &     &  19.8 & (1) \\
      Steph 545A & 05 06 49.6 & $-$21 35 06 & M3.5 &     &  -3.62 &     &   5.3 & (1) \\
          Gl 203 & 05 28 00.1 &    09 38 43 & M3.5 & $<$ &  -4.93 & $<$ &   4.0 & (1) \\
    G 097-052.01 & 05 34 15.1 &    10 19 15 & M3.5 & $<$ &  -4.93 & $<$ &   4.0 & (1) \\
       G 097-054 & 05 34 52.1 &    13 52 48 & M3.5 & $<$ &  -4.93 & $<$ &   2.5 & (2) \\
        LHS 1805 & 06 01 11.1 &    59 35 57 & M3.5 &     &     -- & $<$ &   2.5 & (2) \\
          Gl 273 & 07 27 30.8 &    05 13 12 & M3.5 &     &     -- & $<$ &   2.5 & (2) \\
         Gl 277B & 07 31 57.4 &    36 13 48 & M3.5 &     &  -3.76 & $<$ &   4.0 & (1) \\
         GJ 1105 & 07 58 12.5 &    41 18 19 & M3.5 &     &     -- & $<$ &   2.0 & (5) \\
          Gl 317 & 08 40 59.5 & $-$23 27 31 & M3.5 &     &     -- & $<$ &   2.5 & (2) \\
         GJ 1125 & 09 30 44.8 &    00 19 25 & M3.5 &     &     -- & $<$ &   2.5 & (2) \\
        LHS 2181 & 09 43 55.8 &    26 58 08 & M3.5 & $<$ &  -4.93 & $<$ &   4.0 & (1) \\
          Gl 431 & 11 31 48.0 & $-$41 02 52 & M3.5 &     &  -3.62 &     &  20.5 & (1) \\
          Gl 445 & 11 47 39.2 &    78 41 24 & M3.5 &     &     -- & $<$ &   2.5 & (2) \\
        LHS 2520 & 12 10 05.5 & $-$15 04 10 & M3.5 &     &     -- & $<$ &   2.0 & (5) \\
          Gl 486 & 12 47 57.2 &    09 45 09 & M3.5 &     &     -- & $<$ &   2.5 & (2) \\
        LHS 2651 & 12 55 56.5 &    50 55 28 & M3.5 & $<$ &  -4.93 &     &   3.9 & (6) \\
        LHS 2784 & 13 42 43.2 &    33 17 29 & M3.5 & $<$ &  -4.93 & $<$ &   4.0 & (1) \\
        LHS 2794 & 13 45 50.9 & $-$17 57 56 & M3.5 & $<$ &  -4.93 & $<$ &   2.5 & (2) \\
          Gl 545 & 14 20 07.6 & $-$09 37 07 & M3.5 & $<$ &  -4.93 & $<$ &   4.0 & (1) \\
        Gl 553.1 & 14 31 01.4 & $-$12 17 43 & M3.5 &     &     -- & $<$ &   2.5 & (2) \\
         Gl 568A & 14 53 51.9 &    23 33 18 & M3.5 & $<$ &  -4.93 & $<$ &   4.0 & (1) \\
        LHS 2998 & 14 54 27.7 &    35 33 03 & M3.5 & $<$ &  -4.93 & $<$ &   4.0 & (1) \\
          Gl 628 & 16 30 18.0 & $-$12 39 35 & M3.5 &     &     -- &     &   1.5 & (4) \\
          Gl 643 & 16 55 24.6 & $-$08 19 27 & M3.5 &     &     -- & $<$ &   2.7 & (5) \\
         GJ 1207 & 16 57 05.4 & $-$04 20 52 & M3.5 &     &  -3.80 &     &  10.7 & (1) \\
       G 203-047 & 17 09 31.2 &    43 40 54 & M3.5 & $<$ &  -4.93 & $<$ &   4.0 & (1) \\
       LTT 15087 & 17 11 34.5 &    38 26 33 & M3.5 & $<$ &  -4.93 & $<$ &   2.5 & (2) \\
         Gl 661A & 17 12 07.5 &    45 40 09 & M3.5 & $<$ &  -4.93 & $<$ &   4.0 & (1) \\
         Gl 669A & 17 19 54.5 &    26 30 01 & M3.5 &     &  -4.05 & $<$ &   4.0 & (1) \\
        LHS 3295 & 17 29 25.9 & $-$80 09 07 & M3.5 & $<$ &  -4.93 & $<$ &   3.0 & (1) \\
          Gl 682 & 17 37 04.5 & $-$44 18 57 & M3.5 & $<$ &  -4.93 & $<$ &   3.0 & (1) \\
        LHS 3333 & 17 50 16.0 &    23 45 33 & M3.5 & $<$ & -4.93* & $<$ &   4.5 & (6) \\
       G 205-028 & 18 31 58.4 &    40 41 06 & M3.5 & $<$ &  -4.93 & $<$ &   2.5 & (2) \\
       LP 229-17 & 18 34 36.5 &    40 07 27 & M3.5 &     &     -- & $<$ &   2.5 & (2) \\
         Gl 725B & 18 42 48.0 &    59 37 29 & M3.5 &     &     -- & $<$ &   2.5 & (2) \\
          Gl 729 & 18 49 49.1 & $-$23 50 08 & M3.5 &     &  -3.77 &     &   4.0 & (2) \\
          Gl 748 & 19 12 13.6 &    02 53 15 & M3.5 &     &     -- &     &   4.6 & (3) \\
          Gl 849 & 22 09 45.2 & $-$04 38 11 & M3.5 &     &     -- & $<$ &   2.5 & (2) \\
          Gl 873 & 22 46 50.1 &    44 20 05 & M3.5 &     &  -3.59 &     &   3.5 & (2) \\
        GJ 1005A & 00 15 27.7 & $-$16 07 56 & M4.0 & $<$ &  -4.94 & $<$ &   3.0 & (1) \\
         GJ 1012 & 00 28 39.6 & $-$06 39 44 & M4.0 & $<$ &  -4.94 & $<$ &   3.0 & (1) \\
       LP 525-39 & 00 32 34.7 &    07 29 26 & M4.0 &     &  -3.54 &     &  14.7 & (1) \\
         GJ 1034 & 01 16 30.1 &    24 19 30 & M4.0 & $<$ & -4.94* & $<$ &   4.5 & (6) \\
           Gl 82 & 01 59 23.2 &    58 31 16 & M4.0 &     &  -3.52 &     &  13.8 & (1) \\
        Gl 84.1B & 02 05 24.3 & $-$28 03 20 & M4.0 & $<$ &  -4.94 & $<$ &   3.0 & (1) \\
         Gl 105B & 02 36 14.2 &    06 52 06 & M4.0 &     &     -- & $<$ &   2.5 & (2) \\
        LHS 1426 & 02 37 29.9 &    00 21 26 & M4.0 &     &     -- &     &   4.9 & (6) \\
       Gl 169.1A & 04 31 10.7 &    58 58 53 & M4.0 &     &     -- & $<$ &   2.0 & (5) \\
        LHS 1723 & 05 01 57.6 & $-$06 56 42 & M4.0 &     &  -4.51 & $<$ &   3.2 & (5) \\
        LHS 5109 & 05 35 59.9 & $-$07 39 00 & M4.0 & $<$ &  -4.94 & $<$ &   4.0 & (1) \\
          Gl 213 & 05 42 08.1 &    12 29 33 & M4.0 & $<$ &  -4.94 & $<$ &   2.5 & (2) \\
       G 099-049 & 06 00 03.2 &    02 42 23 & M4.0 &     &  -3.96 &     &   7.4 & (5) \\
          Gl 300 & 08 12 40.8 & $-$21 32 58 & M4.0 & $<$ &  -4.94 & $<$ &   3.0 & (1) \\
        GJ 2069B & 08 31 37.5 &    19 23 48 & M4.0 &     &  -3.85 &     &   6.5 & (5) \\
         Gl 324B & 08 52 41.1 &    28 18 59 & M4.0 &     &     -- & $<$ &   2.5 & (2) \\
       G 234-053 & 09 02 52.6 &    68 03 43 & M4.0 & $<$ &  -4.94 & $<$ &   4.0 & (1) \\
         GJ 1129 & 09 44 48.1 & $-$18 12 47 & M4.0 & $<$ &  -4.94 & $<$ &   3.0 & (1) \\
          Gl 402 & 10 50 52.5 &    06 48 34 & M4.0 &     &     -- & $<$ &   2.5 & (2) \\
         GJ 1148 & 11 41 44.4 &    42 44 00 & M4.0 &     &     -- & $<$ &   2.5 & (2) \\
          Gl 447 & 11 47 44.0 &    00 48 24 & M4.0 &     &     -- & $<$ &   2.5 & (2) \\
       G 165-008 & 13 31 46.7 &    29 16 36 & M4.0 &     &  -3.49 &     &  55.5 & (5) \\
        LHS 2836 & 13 59 10.9 & $-$19 49 59 & M4.0 &     &  -3.69 & $<$ &   3.0 & (1) \\
          Gl 555 & 14 34 17.0 & $-$12 31 15 & M4.0 &     &     -- & $<$ &   2.5 & (2) \\
        LHS 3056 & 15 19 12.1 & $-$12 45 03 & M4.0 & $<$ &  -4.94 & $<$ &   3.0 & (1) \\
          Gl 592 & 15 36 58.8 & $-$14 07 55 & M4.0 & $<$ &  -4.94 & $<$ &   3.0 & (1) \\
          Gl 609 & 16 02 51.4 &    20 35 31 & M4.0 & $<$ &  -4.94 & $<$ &   3.0 & (1) \\
       LP 275-68 & 16 35 27.2 &    35 00 57 & M4.0 &     &  -3.59 &     &  15.8 & (1) \\
          Gl 699 & 17 57 48.1 &    04 43 14 & M4.0 &     &     -- & $<$ &   2.5 & (2) \\
         GJ 1243 & 19 51 09.1 &    46 28 57 & M4.0 &     &  -3.57 &     &  22.0 & (1) \\
         GJ 1254 & 20 33 39.7 &    61 45 07 & M4.0 & $<$ &  -4.94 & $<$ &   4.0 & (1) \\
         Gl 799B & 20 41 50.5 & $-$32 26 00 & M4.0 &     &  -3.07 &     &   8.7 & (1) \\
         GJ 1263 & 21 46 39.9 &    00 10 19 & M4.0 & $<$ &  -4.94 & $<$ &   4.0 & (1) \\
       G 188-038 & 22 01 12.9 &    28 18 24 & M4.0 &     &  -3.52 &     &  35.1 & (1) \\
       G 232-070 & 22 25 16.9 &    59 24 51 & M4.0 & $<$ &  -4.94 & $<$ &   4.0 & (1) \\
         Gl 860B & 22 28 00.3 &    57 41 48 & M4.0 &     &  -4.17 &     &   4.7 & (5) \\
          Gl 876 & 22 53 16.1 & $-$14 15 42 & M4.0 &     &     -- & $<$ &   2.5 & (2) \\
         LHS 543 & 23 21 37.7 &    17 17 35 & M4.0 & $<$ &  -4.94 & $<$ &   2.5 & (2) \\
       G 190-027 & 23 29 22.3 &    41 27 51 & M4.0 &     &  -3.67 &     &  14.5 & (1) \\
         GJ 1289 & 23 43 05.6 &    36 32 13 & M4.0 &     &  -4.56 & $<$ &   2.6 & (5) \\
       G 273-185 & 23 57 19.1 & $-$12 58 40 & M4.0 & $<$ &  -4.94 & $<$ &   4.0 & (1) \\
         Gl 54.1 & 01 12 29.9 & $-$17 00 01 & M4.5 &     &  -4.07 & $<$ &   2.5 & (2) \\
         Gl 83.1 & 02 00 12.3 &    13 03 21 & M4.5 &     &  -4.16 & $<$ &   2.5 & (2) \\
       G 073-045 & 02 20 46.1 &    02 58 39 & M4.5 &     &     -- &     &  23.3 & (6) \\
       G 006-007 & 03 26 45.6 &    19 14 38 & M4.5 &     &     -- &     &  21.9 & (6) \\
         Gl 166C & 04 15 22.9 & $-$07 39 38 & M4.5 &     &  -3.95 &     &   5.5 & (5) \\
         GJ 1078 & 05 23 49.2 &    22 32 40 & M4.5 &     &     -- &     &   7.2 & (6) \\
        LHS 1785 & 05 47 10.2 & $-$05 12 00 & M4.5 & $<$ &  -4.97 &     &   4.5 & (6) \\
          Gl 232 & 06 24 40.9 &    23 26 02 & M4.5 &     &     -- & $<$ &   3.1 & (5) \\
         Gl 234A & 06 29 23.0 & $-$02 48 45 & M4.5 &     &  -3.73 &     &   5.4 & (1) \\
        LHS 1857 & 06 36 07.5 &    11 36 33 & M4.5 & $<$ &  -4.97 &     &   4.0 & (6) \\
        LHS 1885 & 06 57 56.7 &    62 19 23 & M4.5 &     &  -4.20 & $<$ &   3.7 & (5) \\
       Gl 268.01 & 07 10 07.8 &    38 31 27 & M4.5 &     &  -4.11 &     &   6.3 & (1) \\
          Gl 285 & 07 44 40.2 &    03 33 10 & M4.5 &     &  -3.48 &     &   4.5 & (4) \\
        LHS 1950 & 07 51 52.6 &    05 33 28 & M4.5 &     &  -4.97 & $<$ &   2.5 & (6) \\
          Gl 299 & 08 11 57.2 &    08 46 53 & M4.5 & $<$ &  -4.75 &     &   3.0 & (5) \\
       G 194-043 & 08 50 50.7 &    52 53 47 & M4.5 &     &     -- &     &  13.1 & (6) \\
         GJ 1119 & 09 00 32.8 &    46 35 15 & M4.5 &     &  -4.13 & $<$ &   4.0 & (1) \\
         GJ 3542 & 09 17 46.0 &    58 25 22 & M4.5 & $<$ &  -4.97 &     &   3.9 & (6) \\
        LHS 2206 & 09 53 55.4 &    20 56 42 & M4.5 &     &  -3.93 &     &  16.5 & (6) \\
         GJ 1134 & 10 41 38.7 &    37 36 40 & M4.5 & $<$ & -4.97* &     &   4.1 & (6) \\
         GJ 1138 & 10 49 45.8 &    35 32 56 & M4.5 & $<$ &  -4.97 & $<$ &   4.0 & (1) \\
         GJ 1151 & 11 50 58.7 &    48 22 44 & M4.5 &     &     -- & $<$ &   4.1 & (5) \\
       G 121-028 & 11 52 58.1 &    24 28 44 & M4.5 & $<$ & -4.97* &     &   3.8 & (6) \\
         LHS 337 & 12 38 50.4 & $-$38 22 21 & M4.5 & $<$ &  -4.97 & $<$ &   3.0 & (1) \\
        Gl 493.1 & 13 00 33.9 &    05 41 06 & M4.5 &     &  -3.99 &     &  16.8 & (5) \\
         GJ 1186 & 14 53 40.4 &    11 34 25 & M4.5 & $<$ &  -4.97 &     &   3.9 & (6) \\
          Gl 585 & 15 23 51.3 &    17 28 06 & M4.5 & $<$ & -4.97* &     &   3.1 & (6) \\
        LHS 3075 & 15 29 43.5 &    42 52 53 & M4.5 &     &  -4.97 & $<$ &   2.5 & (6) \\
       G 180-011 & 15 55 31.9 &    35 12 00 & M4.5 &     &     -- &     &  21.9 & (6) \\
         Gl 669B & 17 19 53.1 &    26 29 59 & M4.5 &     &  -3.20 &     &   6.1 & (1) \\
         GJ 1224 & 18 07 33.2 & $-$15 57 46 & M4.5 &     &  -3.97 & $<$ &   3.0 & (7) \\
        LHS 3376 & 18 18 56.6 &    66 11 36 & M4.5 &     &  -4.14 &     &  14.6 & (5) \\
         GJ 1227 & 18 22 28.1 &    62 03 10 & M4.5 &     &     -- & $<$ &   2.3 & (5) \\
        LHS 3459 & 19 22 40.9 &    29 26 11 & M4.5 &     &  -4.97 & $<$ &   4.5 & (6) \\
         GJ 1250 & 20 08 21.5 &    33 17 35 & M4.5 &     &     -- &     &  15.7 & (6) \\
        Gl 791.2 & 20 29 47.9 &    09 41 18 & M4.5 &     &  -3.93 &     &  32.0 & (5) \\
         Gl 799A & 20 41 50.5 & $-$32 26 00 & M4.5 &     &  -3.34 &     &  13.6 & (1) \\
         GJ 1268 & 22 24 56.0 &    52 00 27 & M4.5 &     &  -4.97 & $<$ &   4.5 & (6) \\
         Gl 896B & 23 31 51.8 &    19 56 14 & M4.5 &     &  -3.86 &     &  24.2 & (5) \\

%% file: vsinicomp_table.tex
          Gl 424 & M0.0 &     --  & $<$ 2.9 &     --  &     --  &     --  &     --  & $<$ 2.5 & $<$ 2.5 &  \\
       Gl 678.1A & M0.0 & $<$ 4.0 &     --  &     --  &     --  &     --  &     --  & $<$ 2.5 & $<$ 2.5 &  \\
         Gl 720A & M0.0 & $<$ 4.0 &     --  &     --  & $<$ 3.0 &     --  &     --  &     --  & $<$ 3.0 &  \\
          Gl 846 & M0.0 & $<$ 4.0 &     --  &     --  &     --  &     --  &     --  & $<$ 2.5 & $<$ 2.5 &  \\
         Gl 27.1 & M0.5 & $<$ 3.0 &     --  &     --  &     --  &     --  &     --  & $<$ 3.0 & $<$ 3.0 &  \\
          Gl 212 & M0.5 & $<$ 4.0 &     --  &     --  &     --  &     --  &     --  & $<$ 2.5 & $<$ 2.5 &  \\
          Gl 229 & M0.5 & $<$ 3.0 &     --  &     --  &     --  &     1.0 &     --  & $<$ 2.5 &     1.0 &  \\
          Gl 369 & M0.5 & $<$ 3.0 &     --  &     --  &     5.0 &     --  &     --  &     --  & $<$ 3.0 &  \\
         Gl 412A & M0.5 &     --  & $<$ 3.0 &     --  &     --  &     --  &     --  & $<$ 2.5 & $<$ 2.5 &  \\
          Gl 494 & M0.5 &    10.7 &     --  &     --  &     --  &     --  &     --  &     9.7 &     9.7 &  \\
          Gl 514 & M0.5 &     --  & $<$ 2.9 &     --  &     --  &     1.5 &     --  & $<$ 2.5 &     1.5 &  \\
          Gl 809 & M0.5 &     --  & $<$ 2.8 &     --  &     --  &     --  &     --  & $<$ 2.5 & $<$ 2.5 &  \\
            Gl 2 & M1.0 & $<$ 4.0 &     --  &     --  &     --  &     --  &     --  & $<$ 2.5 & $<$ 2.5 &  \\
          Gl 15A & M1.0 &     --  & $<$ 2.9 &     --  &     --  &     --  &     --  & $<$ 2.5 & $<$ 2.5 &  \\
          Gl 390 & M1.0 & $<$ 3.0 &     --  &     --  &     --  &     --  &     --  & $<$ 2.5 & $<$ 2.5 &  \\
          Gl 450 & M1.0 &     --  & $<$ 3.3 &     --  &     --  &     --  &     --  & $<$ 2.5 & $<$ 2.5 &  \\
          Gl 536 & M1.0 & $<$ 4.0 &     --  &     --  &     --  &     --  &     --  & $<$ 2.5 & $<$ 2.5 &  \\
         Gl 570B & M1.0 &     --  &     --  &     --  & $<$ 3.0 &     --  &     --  & $<$ 2.5 & $<$ 2.5 &  \\
          Gl 649 & M1.0 & $<$ 4.0 &     --  &     --  & $<$ 3.0 &     --  &     --  & $<$ 2.5 & $<$ 2.5 &  \\
          Gl 686 & M1.0 &     --  & $<$ 5.0 &     --  &     --  &     --  &     --  & $<$ 2.5 & $<$ 2.5 &  \\
          Gl 701 & M1.0 &     --  & $<$ 3.5 &     --  &     --  &     --  &     --  & $<$ 2.5 & $<$ 2.5 &  \\
          Gl 821 & M1.0 & $<$ 4.0 &     --  &     --  &     --  &     --  &     --  & $<$ 2.5 & $<$ 2.5 &  \\
          Gl 908 & M1.0 &     --  & $<$ 3.0 &     --  &     --  &     --  &     --  & $<$ 2.5 & $<$ 2.5 &  \\
            Gl 1 & M1.5 & $<$ 3.0 &     --  &     --  &     --  &     --  &     --  & $<$ 2.5 & $<$ 2.5 &  \\
         GJ 1009 & M1.5 & $<$ 3.0 &     --  &     --  &     --  &     --  &     --  & $<$ 2.5 & $<$ 2.5 &  \\
           Gl 87 & M1.5 & $<$ 4.0 &     --  &     --  & $<$ 3.0 &     --  &     --  & $<$ 2.5 & $<$ 2.5 &  \\
          Gl 173 & M1.5 & $<$ 3.0 &     --  &     --  &     --  &     --  &     --  & $<$ 2.5 & $<$ 2.5 &  \\
          Gl 205 & M1.5 & $<$ 3.0 & $<$ 2.9 &     --  &     --  &     1.5 &     --  & $<$ 2.5 &     1.5 &  \\
          Gl 382 & M1.5 &     --  & $<$ 2.9 &     --  &     --  &     1.8 &     --  & $<$ 2.5 &     1.8 &  \\
         Gl 414B & M1.5 &     --  &     --  &     --  & $<$ 3.2 &     --  &     --  & $<$ 2.5 & $<$ 2.5 &  \\
          Gl 526 & M1.5 &     --  & $<$ 2.9 &     --  &     --  &     2.0 &     --  & $<$ 2.5 &     2.0 &  \\
         Wo 9492 & M1.5 & $<$ 4.0 &     --  &     --  &     --  &     --  &     --  & $<$ 2.5 & $<$ 2.5 &  \\
          Gl 625 & M1.5 &     --  & $<$ 3.4 &     --  &     --  &     --  &     --  & $<$ 2.5 & $<$ 2.5 &  \\
         Gl 667C & M1.5 & $<$ 3.0 &     --  &     --  &     --  &     --  &     --  & $<$ 2.5 & $<$ 2.5 &  \\
         Gl 745A & M1.5 &     --  & $<$ 3.0 &     --  &     --  &     --  &     --  & $<$ 2.5 & $<$ 2.5 &  \\
          Gl 806 & M1.5 & $<$ 4.0 &     --  &     --  & $<$ 3.0 &     --  &     --  & $<$ 2.5 & $<$ 2.5 &  \\
          Gl 880 & M1.5 &     --  & $<$ 2.8 &     --  &     --  &     --  &     --  & $<$ 2.5 & $<$ 2.5 &  \\
           Gl 70 & M2.0 & $<$ 3.0 & $<$ 3.0 &     --  &     --  &     --  & $<$ 3.0 & $<$ 2.5 & $<$ 2.5 &  \\
          Gl 180 & M2.0 & $<$ 3.0 &     --  &     --  &     --  &     --  &     --  & $<$ 2.5 & $<$ 2.5 &  \\
          Gl 226 & M2.0 & $<$ 4.0 &     --  &     --  &     --  &     --  &     --  & $<$ 2.5 & $<$ 2.5 &  \\
         GJ 2066 & M2.0 &     --  & $<$ 2.7 &     --  &     --  &     --  &     --  & $<$ 2.5 & $<$ 2.5 &  \\
          Gl 393 & M2.0 &     --  & $<$ 2.9 &     --  &     --  &     1.5 &     --  & $<$ 2.5 &     1.5 &  \\
          Gl 411 & M2.0 &     --  & $<$ 2.9 &     --  &     --  &     --  &     --  & $<$ 2.5 & $<$ 2.5 &  \\
        Gl 413.1 & M2.0 & $<$ 3.0 &     --  &     --  &     --  &     --  &     --  & $<$ 2.5 & $<$ 2.5 &  \\
          Gl 465 & M2.0 & $<$ 3.0 &     --  &     --  &     --  &     --  &     --  & $<$ 2.5 & $<$ 2.5 &  \\
          Gl 552 & M2.0 &     --  &     --  &     --  & $<$ 3.0 &     --  &     --  & $<$ 2.5 & $<$ 2.5 &  \\
         Gl 569A & M2.0 & $<$ 4.0 &     --  &     --  & $<$ 3.8 &     --  &     --  & $<$ 2.5 & $<$ 2.5 &  \\
         Gl 745B & M2.0 &     --  &     2.8 &     --  &     --  &     --  &     --  & $<$ 2.5 & $<$ 2.5 &  \\
          Gl 851 & M2.0 &     --  &     --  &     --  & $<$ 3.0 &     --  &     --  & $<$ 2.5 & $<$ 2.5 &  \\
           Gl 26 & M2.5 & $<$ 4.0 &     --  &     --  & $<$ 3.0 &     --  &     --  & $<$ 2.5 & $<$ 2.5 &  \\
          Gl 408 & M2.5 &     --  & $<$ 2.3 &     --  &     --  &     --  &     --  & $<$ 2.5 & $<$ 2.5 &  \\
          Gl 436 & M2.5 & $<$ 4.0 &     --  &     --  & $<$ 3.0 &     --  &     --  & $<$ 2.5 & $<$ 2.5 &  \\
          Gl 694 & M2.5 &     --  &     --  &     --  & $<$ 3.0 &     --  &     --  & $<$ 2.5 & $<$ 2.5 &  \\
         Gl 752A & M2.5 &     --  & $<$ 2.6 &     --  &     --  &     --  &     --  & $<$ 2.5 & $<$ 2.5 &  \\
          Gl 793 & M2.5 &     --  & $<$ 3.2 &     --  &     --  &     --  &     --  & $<$ 2.5 & $<$ 2.5 &  \\
           Gl 48 & M3.0 &     --  & $<$ 2.4 &     --  &     --  &     --  &     --  & $<$ 2.5 & $<$ 2.5 &  \\
    G 244-047.01 & M3.0 & $<$ 4.0 &     --  &     3.2 &     --  &     --  &     --  & $<$ 2.5 & $<$ 2.5 &  \\
          Gl 109 & M3.0 &     --  & $<$ 2.8 &     --  &     --  &     --  &     --  & $<$ 2.5 & $<$ 2.5 &  \\
        LHS 1731 & M3.0 & $<$ 3.0 &     --  &     --  &     --  &     --  &     --  & $<$ 2.5 & $<$ 2.5 &  \\
          Gl 251 & M3.0 &     --  & $<$ 2.4 &     --  &     --  &     --  &     --  & $<$ 2.5 & $<$ 2.5 &  \\
        LHS 1935 & M3.0 & $<$ 3.0 &     --  &     --  &     --  &     --  &     --  & $<$ 2.5 & $<$ 2.5 &  \\
          Gl 362 & M3.0 & $<$ 4.0 &     --  &     --  &     --  &     --  &     --  & $<$ 2.5 & $<$ 2.5 &  \\
          Gl 388 & M3.0 &     --  &     6.2 &     --  &     --  &     3.0 &     3.0 &     2.7 &     3.0 &  \\
          Gl 581 & M3.0 &     --  & $<$ 2.1 &     --  &     --  &     --  &     --  & $<$ 2.5 & $<$ 2.5 &  \\
         Gl 617B & M3.0 & $<$ 4.0 &     --  &     --  &     --  &     --  &     --  & $<$ 2.5 & $<$ 2.5 &  \\
          Gl 655 & M3.0 & $<$ 4.0 &     --  &     --  &     --  &     --  &     --  & $<$ 2.5 & $<$ 2.5 &  \\
          Gl 687 & M3.0 &     --  & $<$ 2.8 &     --  &     --  &     --  &     --  & $<$ 2.5 & $<$ 2.5 &  \\
         LHS 462 & M3.0 & $<$ 4.0 &     --  &     --  &     --  &     --  &     --  & $<$ 2.5 & $<$ 2.5 &  \\
         Gl 725A & M3.0 &     --  & $<$ 2.8 &     --  &     --  &     --  &     --  & $<$ 2.5 & $<$ 2.5 &  \\
         Gl 860A & M3.0 &     --  & $<$ 3.0 &     --  &     --  &     --  &     --  & $<$ 2.5 & $<$ 2.5 &  \\
          Gl 179 & M3.5 & $<$ 3.0 &     --  &     --  &     --  &     --  &     --  & $<$ 2.5 & $<$ 2.5 &  \\
       G 097-054 & M3.5 & $<$ 4.0 &     --  &     --  &     --  &     --  &     --  & $<$ 2.5 & $<$ 2.5 &  \\
        LHS 1805 & M3.5 &     --  & $<$ 2.7 &     --  &     --  &     --  &     --  & $<$ 2.5 & $<$ 2.5 &  \\
          Gl 273 & M3.5 &     --  & $<$ 2.4 &     --  &     --  &     --  &     --  & $<$ 2.5 & $<$ 2.5 &  \\
          Gl 445 & M3.5 &     --  & $<$ 2.0 &     --  &     --  &     --  &     --  & $<$ 2.5 & $<$ 2.5 &  \\
          Gl 486 & M3.5 &     --  & $<$ 2.0 &     --  &     --  &     --  &     --  & $<$ 2.5 & $<$ 2.5 &  \\
        LHS 2794 & M3.5 & $<$ 3.0 &     --  &     --  &     --  &     --  &     --  & $<$ 2.5 & $<$ 2.5 &  \\
          Gl 628 & M3.5 &     --  & $<$ 2.0 &     --  &     --  &     1.5 &     --  & $<$ 2.5 &     1.5 &  \\
       LTT 15087 & M3.5 & $<$ 4.0 &     --  &     --  &     --  &     --  &     --  & $<$ 2.5 & $<$ 2.5 &  \\
       G 205-028 & M3.5 & $<$ 4.0 &     --  &     --  &     --  &     --  &     --  & $<$ 2.5 & $<$ 2.5 &  \\
       LP 229-17 & M3.5 &     --  & $<$ 2.0 &     --  &     --  &     --  &     --  & $<$ 2.5 & $<$ 2.5 &  \\
         Gl 725B & M3.5 &     --  & $<$ 2.8 &     --  &     --  &     --  &     --  & $<$ 2.5 & $<$ 2.5 &  \\
          Gl 729 & M3.5 & $<$ 3.0 &     --  &     --  &     --  &     --  &     4.0 &     4.0 &     4.0 &  \\
          Gl 849 & M3.5 &     --  & $<$ 2.4 &     --  &     --  &     --  &     --  & $<$ 2.5 & $<$ 2.5 &  \\
          Gl 873 & M3.5 &     --  &     6.9 &     --  &     --  &     --  & $<$ 3.0 &     3.5 &     3.5 &  \\
        GJ 1005A & M4.0 & $<$ 3.0 &     --  &     --  &     --  &     --  & $<$ 3.0 &     --  & $<$ 3.0 &  \\
         Gl 105B & M4.0 &     --  & $<$ 2.4 &     --  &     --  &     --  &     --  & $<$ 2.5 & $<$ 2.5 &  \\
          Gl 213 & M4.0 & $<$ 3.0 & $<$ 2.9 &     --  &     --  &     --  &     --  & $<$ 2.5 & $<$ 2.5 &  \\
          Gl 402 & M4.0 &     --  & $<$ 2.3 &     --  &     --  &     --  &     --  & $<$ 2.5 & $<$ 2.5 &  \\
          Gl 447 & M4.0 &     --  & $<$ 2.0 &     --  &     --  &     --  &     --  & $<$ 2.5 & $<$ 2.5 &  \\
          Gl 555 & M4.0 &     --  &     2.7 &     --  &     --  &     --  &     --  & $<$ 2.5 & $<$ 2.5 &  \\
          Gl 699 & M4.0 &     --  & $<$ 2.8 &     --  &     --  &     --  &     --  & $<$ 2.5 & $<$ 2.5 &  \\
       G 188-038 & M4.0 &    35.1 &    29.4 &     --  &     --  &     --  &     --  &     --  &    35.1 &  \\
          Gl 876 & M4.0 &     --  & $<$ 2.0 &     --  &     --  &     --  & $<$ 3.0 & $<$ 2.5 & $<$ 2.5 &  \\
         LHS 543 & M4.0 & $<$ 4.0 &     --  &     --  &     --  &     --  &     --  & $<$ 2.5 & $<$ 2.5 &  \\
         Gl 54.1 & M4.5 & $<$ 3.0 &     --  &     --  &     --  &     --  &     --  & $<$ 2.5 & $<$ 2.5 &  \\
         Gl 83.1 & M4.5 &     --  &     3.8 &     --  &     --  &     --  &     --  & $<$ 2.5 & $<$ 2.5 &  \\
         Gl 234A & M4.5 &     5.4 &     6.0 &     --  &     --  &     --  &     --  &     --  &     5.4 &  \\
          Gl 285 & M4.5 &     --  &     6.5 &     --  &     --  &     4.5 &     5.0 &     4.6 &     4.5 &  \\
          Gl 299 & M4.5 &     --  &     3.0 &     --  &     --  &     --  & $<$ 3.0 &     --  &     3.0 &  \\
         GJ 1119 & M4.5 & $<$ 4.0 &     --  &     4.0 &     --  &     --  &     --  &     --  & $<$ 4.0 &  \\
         GJ 1224 & M4.5 &     --  & $<$ 5.6 &     --  &     --  &     --  & $<$ 3.0 &     --  & $<$ 3.0 &  \\
         GJ 1227 & M4.5 &     --  & $<$ 2.3 &     --  &     --  &     --  & $<$ 3.0 &     --  & $<$ 2.3 &  \\

%% file: paper.bbl
\begin{thebibliography}{60}
\expandafter\ifx\csname natexlab\endcsname\relax\def\natexlab#1{#1}\fi

\bibitem[{{Alekseev}(1998)}]{1998ARep...42..649A}
{Alekseev}, I.~Y. 1998, Astronomy Reports, 42, 649

\bibitem[{{Baraffe} {et~al.}(1998){Baraffe}, {Chabrier}, {Allard}, \&
  {Hauschildt}}]{1998A&A...337..403B}
{Baraffe}, I., {Chabrier}, G., {Allard}, F., \& {Hauschildt}, P.~H. 1998, \aap,
  337, 403

\bibitem[{{Basri} {et~al.}(2000){Basri}, {Mohanty}, {Allard}, {Hauschildt},
  {Delfosse}, {Mart{\'{\i}}n}, {Forveille}, \& {Goldman}}]{2000ApJ...538..363B}
{Basri}, G., {Mohanty}, S., {Allard}, F., {et~al.} 2000, \apj, 538, 363

\bibitem[{{Benedict} {et~al.}(1998){Benedict}, {McArthur}, {Nelan}, {Story},
  {Whipple}, {Shelus}, {Jefferys}, {Hemenway}, {Franz}, {Wasserman},
  {Duncombe}, {van Altena}, \& {Fredrick}}]{1998AJ....116..429B}
{Benedict}, G.~F., {McArthur}, B., {Nelan}, E., {et~al.} 1998, \aj, 116, 429

\bibitem[{Beuzit {et~al.}(2004)Beuzit, S{\'e}gransan, Forveille, Udry,
  Delfosse, Mayor, Perrier, Hainaut, Roddier, Roddier, \&
  Mart{\'\i}n}]{Beuzit:2004p2678}
Beuzit, J.-L., S{\'e}gransan, D., Forveille, T., {et~al.} 2004, Astronomy and
  Astrophysics, 425, 997

\bibitem[{{Bochanski} {et~al.}(2010){Bochanski}, {Hawley}, {Covey}, {West},
  {Reid}, {Golimowski}, \& {Ivezi{\'c}}}]{2010AJ....139.2679B}
{Bochanski}, J.~J., {Hawley}, S.~L., {Covey}, K.~R., {et~al.} 2010, \aj, 139,
  2679

\bibitem[{Browning(2008)}]{Browning:2008p2191}
Browning, M.~K. 2008, The Astrophysical Journal, 676, 1262

\bibitem[{Browning {et~al.}(2010)Browning, Basri, Marcy, West, \&
  Zhang}]{Browning:2010p2615}
Browning, M.~K., Basri, G., Marcy, G.~W., West, A.~A., \& Zhang, J. 2010, The
  Astronomical Journal, 139, 504

\bibitem[{{Burgasser} {et~al.}(2010){Burgasser}, {Simcoe}, {Bochanski},
  {Saumon}, {Mamajek}, {Cushing}, {Marley}, {McMurtry}, {Pipher}, \&
  {Forrest}}]{2010ApJ...725.1405B}
{Burgasser}, A.~J., {Simcoe}, R.~A., {Bochanski}, J.~J., {et~al.} 2010, \apj,
  725, 1405

\bibitem[{{Burningham} {et~al.}(2011){Burningham}, {Leggett}, {Homeier},
  {Saumon}, {Lucas}, {Pinfield}, {Tinney}, {Allard}, {Marley}, {Jones},
  {Murray}, {Ishii}, {Day-Jones}, {Gomes}, \& {Zhang}}]{2011MNRAS.414.3590B}
{Burningham}, B., {Leggett}, S.~K., {Homeier}, D., {et~al.} 2011, \mnras, 414,
  3590

\bibitem[{Cayrel(1988)}]{Cayrel:1988p2705}
Cayrel, R. 1988, The Impact of Very High S/N Spectroscopy on Stellar Physics:
  Proceedings of the 132nd Symposium of the International Astronomical Union
  held in Paris, 132, 345

\bibitem[{Chabrier \& Baraffe(1997)}]{Chabrier:1997p2277}
Chabrier, G., \& Baraffe, I. 1997, Astronomy and Astrophysics, 327, 1039

\bibitem[{{Contadakis}(1995)}]{1995A&A...300..819C}
{Contadakis}, M.~E. 1995, \aap, 300, 819

\bibitem[{Cram \& Mullan(1979)}]{Cram:1979p2487}
Cram, L.~E., \& Mullan, D.~J. 1979, Astrophysical Journal, 234, 579

\bibitem[{da~Silva {et~al.}(2009)da~Silva, Torres, de~La~Reza, Quast, Melo, \&
  Sterzik}]{daSilva:2009p2685}
da~Silva, L., Torres, C. A.~O., de~La~Reza, R., {et~al.} 2009, Astronomy and
  Astrophysics, 508, 833

\bibitem[{Daemgen {et~al.}(2007)Daemgen, Siegler, Reid, \&
  Close}]{Daemgen:2007p2687}
Daemgen, S., Siegler, N., Reid, I.~N., \& Close, L.~M. 2007, The Astrophysical
  Journal, 654, 558

\bibitem[{Delfosse {et~al.}(1998)Delfosse, Forveille, Perrier, \&
  Mayor}]{Delfosse:1998p79}
Delfosse, X., Forveille, T., Perrier, C., \& Mayor, M. 1998, Astronomy and
  Astrophysics, 331, 581

\bibitem[{{Delfosse} {et~al.}(2000){Delfosse}, {Forveille}, {S{\'e}gransan},
  {Beuzit}, {Udry}, {Perrier}, \& {Mayor}}]{2000A&A...364..217D}
{Delfosse}, X., {Forveille}, T., {S{\'e}gransan}, D., {et~al.} 2000, \aap, 364,
  217

\bibitem[{{Demory} {et~al.}(2009){Demory}, {S{\'e}gransan}, {Forveille},
  {Queloz}, {Beuzit}, {Delfosse}, {di Folco}, {Kervella}, {Le Bouquin},
  {Perrier}, {Benisty}, {Duvert}, {Hofmann}, {Lopez}, \&
  {Petrov}}]{2009A&A...505..205D}
{Demory}, B.-O., {S{\'e}gransan}, D., {Forveille}, T., {et~al.} 2009, \aap,
  505, 205

\bibitem[{Donati {et~al.}(2008)Donati, Morin, Petit, Delfosse, Forveille,
  Auri{\`e}re, Cabanac, Dintrans, Fares, Gastine, Jardine, Ligni{\`e}res,
  Paletou, Velez, \& Th{\'e}ado}]{Donati:2008p2516}
Donati, J.-F., Morin, J., Petit, P., {et~al.} 2008, Monthly Notices of the
  Royal Astronomical Society, 390, 545

\bibitem[{{Engle} {et~al.}(2009){Engle}, {Guinan}, \&
  {Mizusawa}}]{2009AIPC.1135..221E}
{Engle}, S.~G., {Guinan}, E.~F., \& {Mizusawa}, T. 2009, in American Institute
  of Physics Conference Series, Vol. 1135, American Institute of Physics
  Conference Series, ed. {M.~E.~van Steenberg, G.~Sonneborn, H.~W.~Moos, \&
  W.~P.~Blair }, 221--224

\bibitem[{{Fekel} \& {Henry}(2000)}]{2000AJ....120.3265F}
{Fekel}, F.~C., \& {Henry}, G.~W. 2000, \aj, 120, 3265

\bibitem[{{Femen{\'{\i}}a} {et~al.}(2011){Femen{\'{\i}}a}, {Rebolo},
  {P{\'e}rez-Prieto}, {Hildebrandt}, {Labadie}, {P{\'e}rez-Garrido},
  {B{\'e}jar}, {D{\'{\i}}az-S{\'a}nchez}, {Vill{\'o}}, {Oscoz}, {L{\'o}pez},
  {Rodr{\'{\i}}guez}, \& {Piqueras}}]{2011MNRAS.413.1524F}
{Femen{\'{\i}}a}, B., {Rebolo}, R., {P{\'e}rez-Prieto}, J.~A., {et~al.} 2011,
  \mnras, 413, 1524

\bibitem[{{Gizis} {et~al.}(2002){Gizis}, {Reid}, \&
  {Hawley}}]{2002AJ....123.3356G}
{Gizis}, J.~E., {Reid}, I.~N., \& {Hawley}, S.~L. 2002, \aj, 123, 3356

\bibitem[{Goldman {et~al.}(2010)Goldman, Marsat, Henning, Clemens, \&
  Greiner}]{Goldman:2010p3162}
Goldman, B., Marsat, S., Henning, T., Clemens, C., \& Greiner, J. 2010, Monthly
  Notices of the Royal Astronomical Society, 405, 1140

\bibitem[{{Gray}(2005)}]{2005oasp.book.....G}
{Gray}, D.~F. 2005, {The Observation and Analysis of Stellar Photospheres}
  (Cambridge University Press)

\bibitem[{Hauschildt {et~al.}(1999)Hauschildt, Allard, \&
  Baron}]{Hauschildt:1999p3036}
Hauschildt, P.~H., Allard, F., \& Baron, E. 1999, The Astrophysical Journal,
  512, 377

\bibitem[{Hawley {et~al.}(1996)Hawley, Gizis, \& Reid}]{Hawley:1996p1924}
Hawley, S.~L., Gizis, J.~E., \& Reid, I.~N. 1996, Astronomical Journal v.112,
  112, 2799

\bibitem[{{Hilton} {et~al.}(2010){Hilton}, {West}, {Hawley}, \&
  {Kowalski}}]{2010AJ....140.1402H}
{Hilton}, E.~J., {West}, A.~A., {Hawley}, S.~L., \& {Kowalski}, A.~F. 2010,
  \aj, 140, 1402

\bibitem[{{Irwin} {et~al.}(2011){Irwin}, {Berta}, {Burke}, {Charbonneau},
  {Nutzman}, {West}, \& {Falco}}]{2011ApJ...727...56I}
{Irwin}, J., {Berta}, Z.~K., {Burke}, C.~J., {et~al.} 2011, \apj, 727, 56

\bibitem[{Jenkins {et~al.}(2009)Jenkins, Ramsey, Jones, Pavlenko, Gallardo,
  Barnes, \& Pinfield}]{Jenkins:2009p2650}
Jenkins, J.~S., Ramsey, L.~W., Jones, H. R.~A., {et~al.} 2009, The
  Astrophysical Journal, 704, 975

\bibitem[{Kenyon \& Hartmann(1995)}]{Kenyon:1995p2462}
Kenyon, S.~J., \& Hartmann, L. 1995, Astrophysical Journal Supplement v.101,
  101, 117

\bibitem[{{Kiraga} \& {Stepien}(2007)}]{Kiraga:2007p407}
{Kiraga}, M., \& {Stepien}, K. 2007, Acta Astron., 57, 149

\bibitem[{{Kowalski} {et~al.}(2009){Kowalski}, {Hawley}, {Hilton}, {Becker},
  {West}, {Bochanski}, \& {Sesar}}]{2009AJ....138..633K}
{Kowalski}, A.~F., {Hawley}, S.~L., {Hilton}, E.~J., {et~al.} 2009, \aj, 138,
  633

\bibitem[{{Lee} {et~al.}(2010){Lee}, {Berger}, \&
  {Knapp}}]{2010ApJ...708.1482L}
{Lee}, K.-G., {Berger}, E., \& {Knapp}, G.~R. 2010, \apj, 708, 1482

\bibitem[{Marcy \& Chen(1992)}]{Marcy:1992p2101}
Marcy, G.~W., \& Chen, G.~H. 1992, Astrophysical Journal, 390, 550

\bibitem[{{Messina} {et~al.}(2003){Messina}, {Pizzolato}, {Guinan}, \&
  {Rodon{\`o}}}]{2003A&A...410..671M}
{Messina}, S., {Pizzolato}, N., {Guinan}, E.~F., \& {Rodon{\`o}}, M. 2003,
  \aap, 410, 671

\bibitem[{Mohanty \& Basri(2003)}]{Mohanty:2003p83}
Mohanty, S., \& Basri, G. 2003, The Astrophysical Journal, 583, 451

\bibitem[{Morin {et~al.}(2008)Morin, Donati, Petit, Delfosse, Forveille,
  Albert, Auri{\`e}re, Cabanac, Dintrans, Fares, Gastine, Jardine,
  Ligni{\`e}res, Paletou, Velez, \& Th{\'e}ado}]{Morin:2008p2170}
Morin, J., Donati, J.-F., Petit, P., {et~al.} 2008, Monthly Notices of the
  Royal Astronomical Society, 390, 567

\bibitem[{Noyes {et~al.}(1984)Noyes, Hartmann, Baliunas, Duncan, \&
  Vaughan}]{Noyes:1984p48}
Noyes, R.~W., Hartmann, L.~W., Baliunas, S.~L., Duncan, D.~K., \& Vaughan,
  A.~H. 1984, Astrophysical Journal, 279, 763

\bibitem[{Ossendrijver(2003)}]{Ossendrijver:2003p2903}
Ossendrijver, M. 2003, The Astronomy and Astrophysics Review, 11, 287

\bibitem[{Parker(1993)}]{Parker:1993p2881}
Parker, E.~N. 1993, Astrophysical Journal, 408, 707

\bibitem[{Pizzolato {et~al.}(2003)Pizzolato, Maggio, Micela, Sciortino, \&
  Ventura}]{Pizzolato:2003p74}
Pizzolato, N., Maggio, A., Micela, G., Sciortino, S., \& Ventura, P. 2003,
  Astronomy and Astrophysics, 397, 147

\bibitem[{{Press} {et~al.}(1992){Press}, {Teukolsky}, {Vetterling}, \&
  {Flannery}}]{1992nrca.book.....P}
{Press}, W.~H., {Teukolsky}, S.~A., {Vetterling}, W.~T., \& {Flannery}, B.~P.
  1992, {Numerical recipes in C. The art of scientific computing}, ed. {Press,
  W.~H., Teukolsky, S.~A., Vetterling, W.~T., \& Flannery, B.~P. }

\bibitem[{Reid {et~al.}(1995)Reid, Hawley, \& Gizis}]{Reid:1995p1938}
Reid, I.~N., Hawley, S.~L., \& Gizis, J.~E. 1995, Astronomical Journal v.110,
  110, 1838

\bibitem[{Reiners(2007)}]{Reiners:2007p18}
Reiners, A. 2007, Astronomy and Astrophysics, 467, 259

\bibitem[{Reiners \& Basri(2007)}]{Reiners:2007p19}
Reiners, A., \& Basri, G. 2007, The Astrophysical Journal, 656, 1121

\bibitem[{{Reiners} \& {Basri}(2008)}]{2008A&A...489L..45R}
{Reiners}, A., \& {Basri}, G. 2008, \aap, 489, L45

\bibitem[{{Reiners} \& {Basri}(2010)}]{2010ApJ...710..924R}
---. 2010, \apj, 710, 924

\bibitem[{{Reiners} \& {Mohanty}(2011)}]{ReiMo11}
{Reiners}, A., \& {Mohanty}, S. 2011, ArXiv e-prints

\bibitem[{Shkolnik {et~al.}(2009)Shkolnik, Liu, \& Reid}]{Shkolnik:2009p2686}
Shkolnik, E., Liu, M.~C., \& Reid, I.~N. 2009, The Astrophysical Journal, 699,
  649

\bibitem[{{Shulyak} {et~al.}(2011){Shulyak}, {Seifahrt}, {Reiners},
  {Kochukhov}, \& {Piskunov}}]{2011MNRAS.tmp.1579S}
{Shulyak}, D., {Seifahrt}, A., {Reiners}, A., {Kochukhov}, O., \& {Piskunov},
  N. 2011, \mnras, 1579

\bibitem[{{Silvestri} {et~al.}(2005){Silvestri}, {Hawley}, \&
  {Oswalt}}]{2005AJ....129.2428S}
{Silvestri}, N.~M., {Hawley}, S.~L., \& {Oswalt}, T.~D. 2005, \aj, 129, 2428

\bibitem[{{Simon} {et~al.}(2006){Simon}, {Bender}, \&
  {Prato}}]{2006ApJ...644.1183S}
{Simon}, M., {Bender}, C., \& {Prato}, L. 2006, \apj, 644, 1183

\bibitem[{Tonry \& Davis(1979)}]{Tonry:1979p1685}
Tonry, J., \& Davis, M. 1979, Astronomical Journal, 84, 1511

\bibitem[{{Torres} {et~al.}(2006){Torres}, {Quast}, {da Silva}, {de La Reza},
  {Melo}, \& {Sterzik}}]{2006A&A...460..695T}
{Torres}, C.~A.~O., {Quast}, G.~R., {da Silva}, L., {et~al.} 2006, \aap, 460,
  695

\bibitem[{{West} \& {Basri}(2009)}]{2009ApJ...693.1283W}
{West}, A.~A., \& {Basri}, G. 2009, \apj, 693, 1283

\bibitem[{{West} {et~al.}(2008){West}, {Hawley}, {Bochanski}, {Covey}, {Reid},
  {Dhital}, {Hilton}, \& {Masuda}}]{2008AJ....135..785W}
{West}, A.~A., {Hawley}, S.~L., {Bochanski}, J.~J., {et~al.} 2008, \aj, 135,
  785

\bibitem[{West {et~al.}(2004)West, Hawley, Walkowicz, Covey, Silvestri,
  Raymond, Harris, Munn, McGehee, Ivezi{\'c}, \& Brinkmann}]{West:2004p78}
West, A.~A., Hawley, S.~L., Walkowicz, L.~M., {et~al.} 2004, The Astronomical
  Journal, 128, 426

\bibitem[{{Zuckerman} {et~al.}(2006){Zuckerman}, {Bessell}, {Song}, \&
  {Kim}}]{2006ApJ...649L.115Z}
{Zuckerman}, B., {Bessell}, M.~S., {Song}, I., \& {Kim}, S. 2006, \apjl, 649,
  L115

\end{thebibliography}
